\documentclass[aps,showpacs,eqsecnum,floatfix]{revtex4}

\usepackage[dvips]{graphicx,color}

\begin{document}

\title{Poincar\'e Invariant Three-Body Scattering at Intermediate
Energies}

\author{T.~Lin$^{(a)}$, Ch.~Elster$^{(a,b)}$}

\author{W.~N.~Polyzou$^{(c)}$}

\author{H.~Wita\l a$^{(d)}$}

\author{W.~Gl\"ockle$^{(e)}$}

\affiliation{(a)
Institute of Nuclear and Particle Physics,  and
Department of Physics and Astronomy,  Ohio University, Athens, OH 45701,
USA}
\affiliation{(b)
Physics Division, Argonne National Laboratory, Argonne, IL 60439, USA}
\affiliation{(c)
Department of Physics and Astronomy, The University of Iowa, Iowa City,
IA 52242, USA}
\affiliation{(d) M. Smoluchowski Institute of Physics, Jagiellonian University, PL-30059
Krak\'ow, Poland}
\affiliation{(e)
Institute for Theoretical Physics II, Ruhr-University Bochum,
D-44780 Bochum, Germany}

\vspace{10mm}

\date{\today}

\vspace{5mm}

\begin{abstract}

The relativistic Faddeev equation for three-nucleon scattering is
formulated in momentum space and directly solved in terms of momentum
vectors without employing a partial wave decomposition.  The equation
is solved through Pad\'e summation, and the numerical
feasibility and stability of the solution is demonstrated.
Relativistic invariance is achieved by constructing a dynamical
unitary representation of the Poincar\'e group on the three-nucleon
Hilbert space.  Based on a Malfliet-Tjon type interaction, observables
for elastic and break-up scattering are calculated for projectile
energies in the intermediate energy range up to 2 GeV, and compared to
their nonrelativistic counterparts. The convergence of the multiple
scattering series is investigated as a function of the 
projectile energy in different scattering observables and configurations. 
Approximations to the two-body interaction embedded in the three-particle 
space are compared to the exact treatment.

\end{abstract}
\vspace{10mm}


\maketitle



\section{Introduction}

The lightest nuclei can be accurately modeled as systems of nucleons
interacting via effective two- and three-body forces motivated e.g. by
meson exchange. This picture is expected to break down at a higher
energy scale, where the physics is more efficiently described in terms
of subnuclear degrees of freedom.  Few-body methods have been an
essential tool for determining model Hamiltonians that describe
low-energy nuclear physics. They also have the potential to be a
useful framework for testing the limitations of viewing the nucleus as
a few nucleon system. The latter requires extending
few-body models and calculations to higher energies. In order to
successfully do this, a number of challenges need to be
addressed. These include replacing the nonrelativistic theory with a
relativistic one, overcoming limitations imposed by interactions fit
to elastic nucleon-nucleon (NN) scattering data, 
including new degrees of freedom that appear above the pion
production threshold, as well as numerical problems related to the
proliferation of partial waves characteristic for scattering
calculations at higher energies. Thus the intermediate energy regime
is a new territory for few-body calculations which waits to be
explored.

In this paper we address two of the challenges. We demonstrate that it
is now possible to perform converged three-body scattering
calculations at energies up to 2~GeV laboratory kinetic energy. Key
elements are a consistent implementation of a Poincar\'e symmetric
quantum theory \cite{Wigner39}, and the use of direct integration
methods that avoid the  partial wave decomposition, 
successfully applied below the pion-production
threshold~\cite{wgphysrep}.  In a series of
publications~\cite{Hang1,Hang2,bound3d} the technique of solving the
nonrelativistic momentum-space Faddeev equation without partial waves
has been mastered, for bound as well as scattering states.  The
relativistic Faddeev equation, based on a Poincar{\'e} invariant mass
operator, has been formulated in detail in~\cite{Lin:2007ck}, showing
that it has both kinematical
and dynamical differences with respect to the corresponding
nonrelativistic equation.

The formulation of the theory is given in a representation of
Poincar\'e invariant quantum mechanics where the interactions are
invariant with respect to kinematic translations and rotations
\cite{Coester65}.  The model Hilbert space is a three-nucleon Hilbert
space (thus not allowing for absorptive processes).  The method used
to introduce the NN interactions in the unitary representation of the
Poincar\'e group allows input of high-precision NN
interactions~\cite{AV18,CDBONN,NIJM} in a way that reproduces the
measured two-body observables. However in this study we use a simpler,
spin-independent interaction consisting of a superposition of an
attractive and a repulsive Yukawa interaction that supports a bound
state with the deuteron binding energy. This is mathematically
equivalent to three-boson scattering.  Poincar\'e invariance and
$S$-matrix cluster properties dictate how the two-body interactions
must be embedded in the three-body dynamical generators.  Scattering
observables are calculated using the Faddeev
equation formulated with the mass Casimir operator (rest
Hamiltonian) constructed from these generators.  We want to point out
that the relativistic Faddeev equation with realistic
spin-dependent interactions has been solved below the pion-production
threshold in  a partial wave basis \cite{Witala1,Witala2,Witala3}.

In order to estimate the size of relativistic effects the interactions
employed in the nonrelativistic and relativistic calculations
presented here are chosen to be two-body phase shift
equivalent.  This is achieved in this article by adding the
interaction to the square of the mass
operator~\cite{cps,Keister:2005eq}.  Thus differences in relativistic
and nonrelativistic calculations first appear in the three-body
calculations.  Those differences are in the choice of kinematic
variables (Jacobi momenta are constructed using Lorentz boosts rather
than Galilean boosts) and in the embedding of the two-body
interactions in the three-body problem, which is a consequence of the
non-linear relation between the two and three-body mass operators.
These differences modify the permutation operators and the off-shell
properties of the kernel of the Faddeev equation.

This article is organized as follows. In Section II the formulation of
the Poincar{\'e} invariant Faddeev equation is given and numerical
aspects for computing the Faddeev kernel are discussed. In Sections
III and IV we present calculations for elastic and breakup processes
in the intermediate energy regime from 0.2 to 1.5~GeV. Our focus here
is the investigation of the convergence of the multiple scattering
series as a function of projectile kinetic energy. We
compare our calculations to selected breakup observables and
investigate a simple approximation of the embedding
of the two-body interaction into the three-body problem.

\section{Solving the Relativistic Faddeev Equation}

A detailed formulation of Poincar{\`e} invariant three-body scattering
has been given in~\cite{Lin:2007ck}, where the driving term in the
relativistic Faddeev equation (first order in the two-body transition
operator) has been used to evaluate cross sections for elastic as well
as break-up scattering.  This is now being complemented by fully
solving the relativistic Faddeev equation based on the numerical
techniques previously used to solve the non-relativistic Faddeev
equation~\cite{Hang1}.  For the convenience of the reader essential
equations are repeated, but for the detailed derivation of the
expressions we refer to Ref.~\cite{Lin:2007ck}.

The symmetrized transition operators $U(z)$ for elastic
scattering and $U_0(z)$ for breakup reactions can be expressed in
terms of the solution $T(z)$ of the symmetrized Faddeev equations
\begin{eqnarray}
U(z) &=& P (z-M_0) + PT(z) \nonumber \\
U_0(z) &=& (1+P)\; T(z) ,
\label{eq:2.1}
\end{eqnarray}
where $M_0$ is the invariant mass operator for three 
non-interacting particles and 
the permutation operator $P$ is given by  $P=P_{12}P_{23}+P_{13}P_{23}$.
The operator $T(z)$ is the solution to the symmetrized 
Faddeev equation
\begin{equation}
T(z)  = T_1(z) P +T_1(z) P (z-M_0)^{-1}  T(z) .
\label{eq:2.2}
\end{equation}
where the operator $T_1(z)$ is the two-body transition operator
embedded in the three-particle Hilbert space and
defined as the solution to
\begin{equation}
T_1(z)=V_1+V_1 (z-M_0)^{-1}T_1(z). 
\label{eq:2.3}
\end{equation}
Here $V_1=V_{23} = M_{23}-M_0$ is the two-body interaction embedded in
the three-body Hilbert space and $M_{23}$ is the
invariant mass operator for two interacting particles and a
spectator.

For calculating transition matrix elements, explicit basis states need
to be introduced.  The momenta of the three particles can be labeled
either by single-particle momenta ${\bf p_1}$, ${\bf p_2}$, and ${\bf
p_3}$, or the total momentum $\mathbf{P}$ and the
relativistic Poincar\'e-Jacobi momenta \cite{Lin:2007ck} 
$\mathbf{q}$ and $\mathbf{k}$.  The explicit relations between the
three-body Poincar\'e Jacobi momenta and the single
particle momenta are
\begin{equation}
\mathbf{q}_i = \mathbf{p}_i + {\mathbf{P} \over M_0}
\left ({\mathbf{P} \cdot \mathbf{p}_i \over M_0 + 
\sqrt{M_0 + \mathbf{P}^2}} - \sqrt{m^2 + \mathbf{p}_i^2} \right ).  
\end{equation}
Then the Poincar\'e-Jacobi momenta $\mathbf{q}$ and $\mathbf{k}$ are
given as
\begin{eqnarray}
{\bf q} & \equiv & {\bf q_i} = - ({\bf q_j} + {\bf q_k} ) \nonumber \\
{\bf k} & \equiv & {\bf k_i} =
{\bf k_{jk}} = \frac{1}{2}\left({\bf
q_j}-{\bf q_k} \right) - \frac{1}{2}\left( {\bf q_j}+{\bf q_k} \right)
\left(\frac{E_j-E_k}{E_j+E_k + \sqrt{(E_j+E_k)^2 -({\bf q_j}+{\bf
q_k})^2}}\right),
\label{eq:2.4}
\end{eqnarray}
where $E_i\equiv E({\bf q_i})=\sqrt{m^2 +{\bf q_i}^2}$.  
In addition, the transformation from the single particle momenta ${\bf
p_i}$ to the Poincar{\'e}-Jacobi momenta has a Jacobian given by
\begin{equation}
|{\bf p_1}, {\bf p_2}, {\bf p_3}\rangle  = 
\left|\frac{\partial({\bf P},{\bf k},{\bf q})}
{\partial({\bf p_1},{\bf p_2},{\bf p_3})}\right|^{1/2}
|{\bf P}, {\bf k}, {\bf q}\rangle 
\label{eq:2.5}
\end{equation}
where for ${\bf P}=\mathbf{0}$ the Jacobian becomes
\begin{equation}
\left|\frac{\partial({\bf P},{\bf k},{\bf q})} {\partial({\bf
p_1},{\bf p_2},{\bf p_3})}\right|^{1/2}_{\vert_{{\bf P}=\mathbf{0}}} = \left(
\frac{\sqrt{(E({\bf
q_2})+E({\bf q_3}))^2 -{\bf q}^2} \;(E({\bf q_2})+E({\bf q_3}))}
{4E({\bf q_2})E({\bf q_3})} \right) ^{1/2} .
\label{eq:2.6}
\end{equation}

In the above expression we chose, without loss of
generality, particle 1 as the spectator.  The
Poincar{\'e}-Jacobi momenta are relevant for the calculation of the
permutation operator $P$ in Eqs.~(\ref{eq:2.1}) and (\ref{eq:2.2}).
The matrix elements of the permutation operator are then explicitly
calculated as
\begin{eqnarray}
\langle {\bf k'},{\bf q'}|P|{\bf k},{\bf q}\rangle &=& N({\mathbf
q}',{\mathbf q}) \Bigg[ \delta ({\mathbf k}'-{\mathbf q} -
\frac{1}{2}{\mathbf q}' \: C({\mathbf q},{\mathbf q}') )
\delta({\mathbf k}+{\mathbf q}' +\frac{1}{2}{\bf q} \: C(({\bf
q'},{\bf q})) \nonumber \\ &+& \delta({\mathbf k}'+{\mathbf q}
+\frac{1}{2}{\bf q}' \: C({\bf q},{\bf q}')) \delta({\mathbf
p}-{\mathbf q}' -\frac{1}{2}{\bf q} \: C(({\bf q'},{\bf q}))\Bigg],
\label{eq:2.7}
\end{eqnarray}
where the function $N({\mathbf q}',{\mathbf q})$ contains the
product of two Jacobians and reads
\begin{eqnarray}
N(\mathbf{q},\mathbf{q}') \equiv N(q,q',x') &=& \frac{\sqrt{E(\mathbf{q}) +
E(\mathbf{q} + \mathbf{q}')} \; \sqrt{E(\mathbf{q}') + E(\mathbf{q} +
\mathbf{q}')}} {4E(\mathbf{q} + \mathbf{q}') } \nonumber\\ 
& & \times
\frac{\sqrt[4]{(E(\mathbf{q}) + E(\mathbf{q} + \mathbf{q}'))^2 -
\mathbf{q}'^2} \;\; \sqrt[4]{(E(\mathbf{q}') + E(\mathbf{q} + \mathbf{q}'))^2
- \mathbf{q}^2}} {\sqrt{E(\mathbf{q})E(\mathbf{q}')}} .
\label{eq:2.8}
\end{eqnarray}
with $x' = {\bf {\hat q}} \cdot {\bf {\hat q}'}$.
The function $C({\mathbf q},{\mathbf q}')$ is calculated as
\begin{equation}
C({\mathbf q}',{\mathbf q}) \equiv C(q',q,x') 
= 1 + \frac{E({\mathbf q}')-E({\mathbf q}'+{\mathbf q})}
{E({\mathbf q}')+E({\mathbf q}'+{\mathbf q})
+ \sqrt{(E({\mathbf q}')+E({\mathbf q}'+{\mathbf q}))^2 -{\mathbf q}^2} } .
\label{eq:2.9}
\end{equation}
These permutation operators, which change the order of coupling, are
essentially Racah coefficients for the Poincar\'e group.  In the
nonrelativistic case the functions $N({\mathbf q}',{\mathbf q})$ and
$C({\mathbf q}',{\mathbf q})$ both reduce to the constant 1 and have
the relatively compact form of the matrix elements of $P$ given in
e.g. \cite{Hang1,bound3d}.

In matrix form the Faddeev equation, Eq.~(\ref{eq:2.2}), reads
\begin{equation}
\langle \mathbf{k}, \mathbf{q} \Vert T \Vert \varphi_d, \mathbf{q}_0 \rangle =
\langle \mathbf{k}, \mathbf{q} \Vert T_1 P 
\Vert \varphi_d, \mathbf{q}_0 \rangle +
\langle \mathbf{k}, \mathbf{q} \Vert T_1 P(z-M_0)^{-1} 
T \Vert \varphi_d, \mathbf{q}_0
\rangle , 
\label{eq:2.10}
\end{equation}
where we have factored out a delta function in the total
momentum and set $\mathbf{P}=\mathbf{0}$.
Inserting complete sets of states and explicitly evaluating the permutation operator
 leads to
\begin{eqnarray}
\langle \mathbf{k}, \mathbf{q} \Vert T({\sf W}) \Vert \varphi_d, \mathbf{q}_0 \rangle
&=& N(\mathbf{q},\mathbf{q}_0) \; T_s\left(\mathbf{k},\mathbf{q}_0 +
\frac{1}{2}\mathbf{q} \:
C(\mathbf{q}_0,\mathbf{q}),\mathbf{q};\varepsilon \right) \varphi_d
\left(\mathbf{q} + \frac{1}{2}\mathbf{q}_0 \:
C(\mathbf{q},\mathbf{q}_0) \right)  \nonumber \\
&+& \int d^3 q' \: N(\mathbf{q},\mathbf{q}')
 \frac{T_s(\mathbf{k}, \mathbf{q}' +
\frac{1}{2}\mathbf{q}C(\mathbf{q}',\mathbf{q}),\mathbf{q};\varepsilon)
\langle \mathbf{q} + \frac{1}{2}\mathbf{q}'C(\mathbf{q},\mathbf{q}'), \mathbf{q}' \Vert
T ({\sf W}) \Vert  \varphi_d \mathbf{q}_0 \rangle}
{ W-\left(\sqrt{m^2+\mathbf{q}^2}+\sqrt{m^2+\mathbf{q}'^2}
  +\sqrt{m^2+(\mathbf{q}+\mathbf{q}' )^2} \right)+i\epsilon } .
\label{eq:2.11}
\end{eqnarray}
The quantities ${\sf W}$ and $\mathbf{q}_0$ are determined
by the 
laboratory kinetic energy $E_{lab}$ of the incident nucleon,
\begin{equation}
{\sf W}^2 = ( m+m_d )^2 + 2m_d E_{lab}.
\label{eq:2.12}
\end{equation}
The nucleon rest mass is given by $m$, the rest mass of the deuteron
is $m_d = 2m - \varepsilon_d$, where $\varepsilon_d$ is the deuteron
binding energy. The Poincar\'e Jacobi momentum between projectile and
target, ${\bf q_0}$, is related to $E_{lab}$ by
\begin{equation} {\bf q_0^2} = \frac{m_d^2 E_{lab}}{{\sf W}^2} \left(
{E_{lab} + 2m} \right). 
\label{eq:2.13} 
\end{equation} 
The  invariant parametric energy $\varepsilon$ which enters the two-body
t-matrix is given by $\varepsilon = W -\sqrt{m^2+\mathbf{q}^2}$. Since
we consider bosons, we introduce the symmetrized two-body transition matrix $T_s$
\begin{eqnarray}
T_s(\mathbf{k},\mathbf{k}',\mathbf{q};\varepsilon) &=&
T_1(\mathbf{k},\mathbf{k}',\mathbf{q};\varepsilon) +
T_1(-\mathbf{k},\mathbf{k}',\mathbf{q};\varepsilon) \nonumber \\ &=&
T_1(\mathbf{k},\mathbf{k}',\mathbf{q};\varepsilon) + 
T_1(\mathbf{k},-\mathbf{k}',\mathbf{q};\varepsilon) . 
\label{eq:2.14}
\end{eqnarray}
This two-body t-matrix has a simple pole at ${\sf W} = \sqrt{m^2+\mathbf{q}^2}
+\sqrt{m_d^2+\mathbf{q}^2}$. Thus, for the practical calculation we need to 
take this pole explicitly into consideration by defining
\begin{eqnarray}
\hat{T_s} &=& \left({\sf W} - \sqrt{m^2+\mathbf{q}^2} -\sqrt{m_d^2+\mathbf{q}^2} 
\right) \ T_s   \nonumber \\
\hat{T} &=& \left({\sf W} - \sqrt{m^2+\mathbf{q}^2} -\sqrt{m_d^2+\mathbf{q}^2} 
\right) T .
\label{eq:2.15}
\end{eqnarray}
and solving Eq.~({\ref{eq:2.11}}) for ${\hat T}$
\begin{eqnarray}
\lefteqn { \langle \mathbf{k}, \mathbf{q} \Vert \hat{T}(W) \Vert \varphi_d
,\mathbf{q}_0 \rangle } \nonumber \\ 
&=& N(\mathbf{q},\mathbf{q}_0) \hat{T}_s\left(\mathbf{k},\mathbf{q}_0 +
\frac{1}{2}\mathbf{q} \:
C(\mathbf{q}_0,\mathbf{q}),\mathbf{q};\varepsilon \right) \varphi_d
\left(\mathbf{q} + \frac{1}{2}\mathbf{q}_0 \:
C(\mathbf{q},\mathbf{q}_0) \right)  \nonumber \\
& & + \int d^3q' N(\mathbf{q},\mathbf{q}')
 \frac{\hat{T}_s(\mathbf{k}, \mathbf{q}' +
\frac{1}{2}\mathbf{q}C(\mathbf{q}',\mathbf{q}),\mathbf{q};\varepsilon)}
{W - \left(\sqrt{m^2+\mathbf{q}'^2} +\sqrt{m_d^2+\mathbf{q}'^2}\right) +
i\epsilon}
\nonumber \\
& & \times \frac{\langle \mathbf{q} +
\frac{1}{2}\mathbf{q}'C(\mathbf{q},\mathbf{q}'), \mathbf{q}' \Vert  \hat{T}
\Vert  \varphi_d \mathbf{q}_0 \rangle}
{ W-\left(\sqrt{m^2+\mathbf{q}^2}+\sqrt{m^2+\mathbf{q}'^2}
  +\sqrt{m^2+(\mathbf{q}+\mathbf{q}' )^2} \right)+i\epsilon }  .
\label{eq:2.16}
\end{eqnarray}
For the explicit calculation we introduce the independent variables
\cite{Hang1}
\begin{equation}
k=|{\mathbf{k}}|,\  q=|{\mathbf{q}}|,\
x_{k}=\hat{{\mathbf{k}}}\cdot\hat{{\mathbf{q}}}_{0},\
x_{q}=\hat{{\mathbf{q}}}\cdot\hat{{\mathbf{q}}}_{0},\
x^{q_{0}}_{kq}=
\widehat{({\mathbf{q}}_{0}\times{\mathbf{q}})}\cdot
\widehat{({\mathbf{q}}_{0}\times{\mathbf{k}})},
\label{eq:2.17}
\end{equation}
so that $\langle \mathbf{k}, \mathbf{q} | {\hat T} |
\varphi_d ,\mathbf{q}_0 \rangle = {\hat T} (k, x_k, x^{q_0}_{kq},
x_{q}, q)$, is a function of 5 variables.  In the variables of
Eq.~(\ref{eq:2.17}) and defining $y_{kq} = x_kx_q +
\sqrt{1-x_k^2}\sqrt{1-x_q^2} x^{q_0}_{kq}$, the final expression for
Eq.~({\ref{eq:2.11}}) reads
\begin{eqnarray}
\lefteqn {\hat{T}(k, x_k, x^{q_0}_{kq}, x_{q}, q) } \nonumber \\ 
&=& N(q,q_0,x_q) 
\varphi_d \left(|\mathbf{q} + \frac{1}{2}\mathbf{q}_0 \:
C(\mathbf{q},\mathbf{q}_0)| \right) \nonumber \\ 
& & \times
\hat{T}_s \left(k,|\mathbf{q}_0 + \frac{1}{2}\mathbf{q}
\:C(\mathbf{q}_0,\mathbf{q})|, 
y_{k,\mathbf{q}_0 + \frac{1}{2}\mathbf{q} \:C(\mathbf{q}_0,\mathbf{q})} ;
W -\sqrt{m^2+\mathbf{q}^2} \right) \nonumber \\
& & + \int d^3q' N(q,q',x')
 \frac{\hat{T}_s \left(k,|\mathbf{q}' + \frac{1}{2}\mathbf{q}
\:C(\mathbf{q}',\mathbf{q})|, 
y_{k,\mathbf{q}' + \frac{1}{2}\mathbf{q} \:C(\mathbf{q}',\mathbf{q})};
W -\sqrt{m^2+\mathbf{q}^2} \right) }
{W - \left(\sqrt{m^2+\mathbf{q}'^2} +\sqrt{m_d^2+\mathbf{q}'^2}\right) +
i\epsilon}
\nonumber \\
& & \times \frac{\hat{T} \left(|\mathbf{q} +
\frac{1}{2}\mathbf{q}'C(\mathbf{q},\mathbf{q}')|,
y_{\mathbf{q} + \frac{1}{2}\mathbf{q}'C(\mathbf{q},\mathbf{q}'),q_0},
x^{q_0}_{\mathbf{q} + \frac{1}{2}\mathbf{q}'C(\mathbf{q},\mathbf{q}'),q'},
y_{q'q_0}, q' \right) }
{ W-\left(\sqrt{m^2+\mathbf{q}^2}+\sqrt{m^2+\mathbf{q}'^2}
  +\sqrt{m^2 + (q^2+q'^2 + 2qq'x'} \right)+i\epsilon }.
\label{eq:2.19}
\end{eqnarray}
While the deuteron pole can be numerically taken care of with a single
subtraction in the $q'$-integration, the free
three-nucleon propagator  in the 2nd term under the
integral of Eq.~(\ref{eq:2.19}) contains singularities depending on
$q'$ as well as $x'$ leading to a singular region in the $q-q'$ plane.
In order to simplify the calculation, we carry out the integration of
the kernel in a frame in which the $z$-axis is along the direction of
$\mathbf{q}$. In this frame $x'=\mathbf{\hat{q'}} \cdot
\mathbf{\hat{q}}$ and $\phi'$ is the azimuthal angle of
$\mathbf{q}'$. With these definitions one has
\begin{eqnarray}
y_{q'q_0} &=& x_qx' + \sqrt{1-x_q^2}\sqrt{1-x'^2} \; \cos(\phi_{q_0}-\phi')
\nonumber \\
y_{kq'} &=& x_px' + \sqrt{1-x_k^2}\sqrt{1-x'^2} \; \cos(\phi_{k}-\phi') ,
\label{eq:2.20}
\end{eqnarray}
where $\phi_{k}$ and $\phi_{q_0}$ are the azimuthal angles of
$\mathbf{k}$ and $\mathbf{q}_0$ in the frame described above.  Since
there is a freedom in choosing the $x$-axis, we may place
$\mathbf{q}_0$ in the $xz$-plane, this gives $\phi_{q_0}=0$. With this
choice $\phi_{k}$ is evaluated as
\begin{equation}
\cos\phi_{k} = \frac{x_k - y_{kq}x_q}{\sqrt{1-y_{kq}^2}\sqrt{1-x_q^2}}.
\label{2.21}
\end{equation}
The remaining variables in Eq.~(\ref{eq:2.19})  are explicitly evaluated as
\begin{eqnarray}
|\mathbf{q} + \frac{1}{2}\mathbf{q}'\; C(\mathbf{q},\mathbf{q}')| 
 &=& \sqrt{q^2 + \frac{1}{4}q'^2 \; C^2(q,q',x') +
qq'x'\; C(q,q',x')} \nonumber \\
|\mathbf{q}_0 + \frac{1}{2}\mathbf{q} \; C(\mathbf{q}_0,\mathbf{q})|
 &=& \sqrt{q_0^2 + \frac{1}{4}q^2 \; C^2(q_0,q,x_q) +
qq_0x_q \; C(q_0,q,x_q) } \nonumber \\
|\mathbf{q}' + \frac{1}{2}\mathbf{q}\; C(\mathbf{q}',\mathbf{q})|
 &=& \sqrt{q'^2 + \frac{1}{4}q^2 \; C^2(q',q,x') +
qq'x'\; C(q',q,x')} ,
\label{2.22}
\end{eqnarray}
and
\begin{eqnarray}
y_{k,\mathbf{q}_0 + \frac{1}{2}\mathbf{q} \;C(\mathbf{q}_0,\mathbf{q})} 
&=& \frac{\mathbf{k} \cdot (\mathbf{q}_0 + \frac{1}{2}\mathbf{q}
\;C(\mathbf{q}_0,\mathbf{q}))}
{k|\mathbf{q}_0 + \frac{1}{2}\mathbf{q} \;C(\mathbf{q}_0,\mathbf{q})|} \nonumber \\
 &=& \frac{kq_0x_p+\frac{1}{2}kq y_{kq} \; C(q_0,q,x_q)}
{k \sqrt{q_0^2 + \frac{1}{4}q^2 \; C^2(q_0,q,x_q) +
qq_0x_q \; C(q_0,q,x_q)}} \nonumber \\
y_{k,\mathbf{q}' + \frac{1}{2}\mathbf{q} \:C(\mathbf{q}',\mathbf{q})}
&=& \frac{\mathbf{k} \cdot (\mathbf{q}' + \frac{1}{2}\mathbf{q}
\:C(\mathbf{q}',\mathbf{q}))}
{k|\mathbf{q}' + \frac{1}{2}\mathbf{q} \:C(\mathbf{q}',\mathbf{q})|} \nonumber \\
&=& \frac{kq'y_{kq'}+\frac{1}{2}kq y_{kq} \; C(q',q,x')}
{k\sqrt{q'^2 + \frac{1}{4}q^2 \; C^2(q',q,x') +
qq'x'\; C(q',q,x')}} \nonumber \\
y_{\mathbf{q} + \frac{1}{2}\mathbf{q}'C(\mathbf{q},\mathbf{q}'),q_0}
 &=& \frac{(\mathbf{q} + \frac{1}{2}\mathbf{q}'C(\mathbf{q},\mathbf{q}')) \cdot
\mathbf{q}_0}
          {q_0 |\mathbf{q} + \frac{1}{2}\mathbf{q}'C(\mathbf{q},\mathbf{q}')|}
\nonumber \\
&=& \frac{qq_0x_q + \frac{1}{2}q'q_0 y_{q_0q'} \; C(q,q',x')) }
          {q_0\sqrt{q^2 + \frac{1}{4}q'^2  \; C^2(q,q',x') +
qq'x' \; C(q,q',x')}}, 
\label{2.23}
\end{eqnarray}
and
\begin{eqnarray}
x^{q_0}_{\mathbf{q} + \frac{1}{2}\mathbf{q}'C(\mathbf{q},\mathbf{q}'),q'}
&=& \frac{y_{\mathbf{q} + \frac{1}{2}\mathbf{q}'C(\mathbf{q},\mathbf{q}'),q'} 
- y_{\mathbf{q} + \frac{1}{2}\mathbf{q}'C(\mathbf{q},\mathbf{q}'),q_0}
  y_{q_0q'} }
    {\sqrt{1-y_{\mathbf{q} +
\frac{1}{2}\mathbf{q}'C(\mathbf{q},\mathbf{q}'),q_0}^2}\sqrt{1-y_{q_0q'}^2}},
\nonumber \\
\end{eqnarray}
with
\begin{eqnarray}
y_{\mathbf{q} + \frac{1}{2}\mathbf{q}'C(\mathbf{q},\mathbf{q}'),q'}
&=& \frac{(\mathbf{q} + \frac{1}{2}\mathbf{q}'C(\mathbf{q},\mathbf{q}')) \cdot
\mathbf{q}'}
          {q' |\mathbf{q} + \frac{1}{2}\mathbf{q}'C(\mathbf{q},\mathbf{q}')|}
\nonumber \\
&=& \frac{qq'x' + \frac{1}{2} q'^2 \; C(q,q',x'))}
    {q' \sqrt{q^2 + \frac{1}{4}q'^2 \; C^2(q,q',x') +
qq'x' \; C(q,q',x')} } .
\end{eqnarray}

For the integration of the 3N propagator, each singularity in the
$x'$-integration (for fixed $q'$) is explicitly taken into account by
a subtraction. However, this leads to logarithmic singularities in
$q'$ at the boundaries $x'=\pm 1$. These we integrate in the
semi-analytic fashion introduced in Ref.~\cite{Hang1} by using cubic
Hermite splines. While using cubic Hermite splines is advantageous in
dealing with the logarithmic singularities, this method is not as
effective as Gauss-Legendre quadrature when integrating over large,
non-singular regions. Thus, in order to make the most efficient use of
both methods, we divide the interval of the $q'$-integration into
several integration regions, and use Gauss-Legendre quadrature in the
non-singular integrals, while keeping the cubic Hermite splines in the
small regions around the singularities. With this procedure we are
able to successfully integrate  over  the Faddeev 
kernel with sufficient
accuracy.  For the final solution of Eq.~(\ref{eq:2.19}) the kernel is
successively applied and the resulting terms are summed up as Pad{\'e}
sums.  At the higher energies we will also carry out the Neumann sum.

Our explicit calculations are based on a simple interaction of
Malfliet-Tjon type consisting of a superposition of an attractive and
repulsive Yukawa interaction that supports a bound state with the
deuteron binding energy. The parameters of this nonrelativistic
interaction are given in Ref.~\cite{Hang1}. In order to obtain a
relativistic interaction which is phase shift equivalent with the
nonrelativistic one, we employ a scheme in which $4m$ multiplied with
the interaction is added to the square of the
non-interacting two-body mass operator. This procedure
was introduced by Coester, Pieper and Serduke \cite{cps} and used here
in the form given in \cite{Keister:2005eq}. It guarantees that
differences in the relativistic and nonrelativistic calculations first
appear in the three-body calculations.

Before entering a detailed study on relativistic effects, we want to
present further details on the numerical quality of our solution of
the relativistic Faddeev equation.  One internal consistency check of
the solution is provided by the optical theorem, which states that the
total cross section, being the sum of the total elastic cross section,
$\sigma_{el}$, and the total breakup cross section, $\sigma_{br}$,
must be equal to the imaginary part of the transition operator for
elastic scattering $U$ in forward direction.  In the 
center-of-momentum 
(c.m.) frame this relation reads
\begin{equation}
\sigma_{el} + \sigma_{br} = \sigma_{tot} =  - 16\pi^3 \frac{ E_n(q_0)E_d(q_0) }{q_0W} 
\ \Im m ( U(q_0,x=1)).
\label{opticaltheorem}
\end{equation}
Listed in Table~\ref{table-1} are our fully relativistic calculations
of the total cross sections for elastic scattering and breakup
reaction for projectile energies from 0.1 to 2.0~GeV, together with
the total cross sections.  The total cross sections are calculated as
sum of the elastic and breakup cross sections, $\sigma_{tot}$, and via
the optical theorem, $\sigma_{op}$, from the imaginary part of the
operator $U$ in forward direction, $x=1$. A comparison of those two
numbers for the total cross section shows, that our calculations
fulfill the optical theorem to about 1\% or better up to 1~GeV. This
error increases to about 3\% at 2~GeV. Here we did not push the
calculations any further, since our model potential is too simple to
take it to much higher energies anyway. For the sake of showing the
numerical quality of our calculations, we included 2~GeV in
Table~\ref{table-1}, but will not show any further observables at this
energy.
  
\begin{figure}
\begin{center}
 \includegraphics[width=8cm]{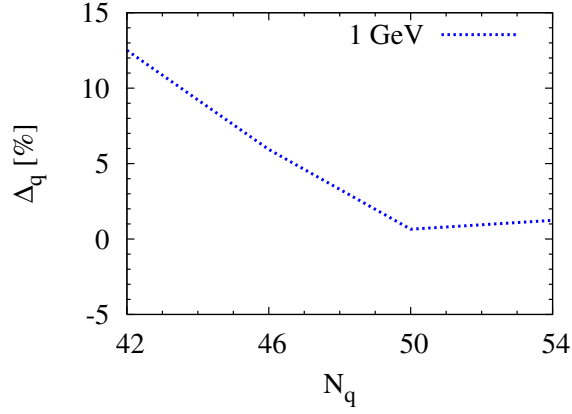}
\end{center}
\caption{(Color online) The percent error $\Delta_q =
\frac{\sigma_{op}-\sigma_{tot}}{\sigma_{op}} \times 100$ 
in the optical theorem as a function of the grid points in the momentum $q$ 
for a calculation at 1.0~GeV.
\label{fig1}}
\end{figure}

\begin{figure}
\begin{center}
 \includegraphics[width=8cm]{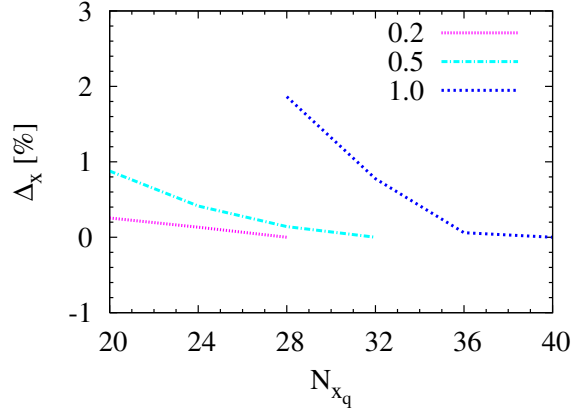}
\end{center}
\caption{(Color online) The percent error $\Delta_{x} =
\frac{\sigma_{op}-\sigma_{tot}}{\sigma_{op}} \times 100$
in the optical theorem as a function of the grid points in the angle grid $x_q$,
when this grid is increased successively by 4 Gauss-Legendre points. The
different curves correspond to the three different laboratory projectile energies
in GeV, indicated in the legend. 
\label{fig2}}
\end{figure}

The transition amplitude of Eq.~(\ref{eq:2.19}) is a function of 5
variables, and is the solution of an integral equation in three
dimensions.  Thus, in the calculation the dependence of the result on
the various choices of grids has to be considered.  As
far as the momentum grids are concerned, the accuracy of
the calculation is most sensitive to the $q$-grid, as already found in
Ref.~\cite{Hang1}.  In Fig.~\ref{fig1} we show the dependence of the
relative error $\Delta_q =
\frac{\sigma_{op}-\sigma_{tot}}{\sigma_{op}}\times 100$ in the optical
theorem as function of the size of the $q$-grid, $N_q$, for a
calculation at 1~GeV projectile laboratory kinetic energy. The slope
of $\Delta_q$ shows that indeed the accuracy of the calculation is
strongly influenced by the size of this grid. For our calculation,
$N_q$~=~50 is sufficient at 1~GeV.  Next, we consider the sensitivity
of the calculation to the size of different angle grids. In
Table~\ref{table-2} we give the cross sections for elastic 
scattering and breakup reactions
together with the total cross section $\sigma_{op}$ extracted from the
optical theorem when varying the size of the 
different angle grids.  We can see,
that the results are most sensitive with respect to the grids in $x_q$
and $x'$.  It is common wisdom in calculations using an angular
momentum basis that as the energy of the projectile increases, the
number of partial waves needed to obtain a converged result increases
rather quickly.  In our 3D calculations, all partial waves are included. The
increase in energy manifests itself in a two-body t-matrix acquiring a
more pronounced peak structure in the forward and
backward directions with respect to the angle between the 
two momentum vectors~\cite{t-matrix}. This peak structure at $x_q=\pm
1$ must be adequately covered in calculations at higher energies to
ensure converged results. In Fig.~\ref{fig2} we show the relative
error $\Delta_x = \frac{\sigma_{op}-\sigma_{tot}}{\sigma_{op}}\times
100$ in the optical theorem as function of the size of the $x_q$-grid for three
different projectile laboratory kinetic energies.  The necessity of
increasing the $x_q$-grid with increasing projectile energy is clearly
seen. Whereas for 0.2~GeV $N_{x_q}$~=~24 is clearly sufficient, at
0.5~GeV one needs already at least 28 points, whereas at 1~GeV a
minimum of 36 points is required. This conclusion is also reached in
our Table~\ref{table-3}, where we show the relativistic differential
cross section at selected angles while varying the $x_q$-grid.  It
should be noted that the angle $x_q$ is related to the angular
momentum of the relative motion between the spectator and the
interacting pair.  The angle $x_p$, which is related to the angular
momentum of the interacting pair is not nearly as sensitive as
$x_q$. In Table~\ref{table-2} we vary the $x_p$ grid from 20 to 24
points, and see hardly any difference.

It is illustrative to contrast the computational algorithm for 
direct integration 
with the experience gained when using a partial wave basis in the 3N
system.  Our experience tells us that at $E_{lab}$~=~200~MeV the total
angular momentum of the 2N subsystem $j$ needs to be $j_{max}$~=~5 to
reach convergence. Furthermore, the maximum total angular momentum $J$
of the 3N system required to reach convergence is
$J_{max}$~=~25/2. Let us assume that $J_{max} = j_{max} + I_{max}$,
where $I_{max}  = s_i + \lambda$ is the maximal angular
momentum of the projectile nucleon with respect to the target
pair, $s_1$ is the spin of the projectile and $\lambda$ is
the relative orbital angular momentum between the projectile and
target pair.  This leads to $I_{max}$~=~15/2 for a 3N scattering
calculation at $E_{lab}$~=~200~MeV. Disregarding the spin degree of
freedom for the three nucleons, leading to the three-boson model under
consideration here, we find that $J_{max}$~=~12 with $l_{max}$~=~5 and
$\lambda_{max}$~=~7 are
necessary for a converged calculation at $E_{lab}$~=~200~MeV.  In the
three-boson case $l$, the orbital momentum of the
interacting pair, and $\lambda$ take the role of $j$ and $I$.

In order to estimate the corresponding maximal number of angular
momenta needed for $E_{lab}$~=~1~GeV, one needs the effective deuteron
radius $r_0$, which leads to $\lambda_{max}$~=~7 at
$E_{lab}$~=~200~MeV.  Nonrelativistically the 3N c.m. energy is given
as $3/4 \; q_0^2 = 2/3 \; E_{lab}$, leading to $q_0 \simeq$~400~MeV/c at
$E_{lab}$~=~200~MeV and $q_0 \simeq$~900~MeV/c at $E_{lab}$~=~1~GeV.
 If we roughly set $\lambda_{max} = q_0 \times r_0$, then
we find at $E_{lab}$~=~200~MeV  a value $r_0 \simeq$~3.5~fm, which appears
reasonable. Applying the same value at $E_{lab}$~=~1~GeV then leads to
$\lambda_{max}$~=~15. Using our experience in calculating the NN
system in the GeV regime~\cite{Pricking:2007qx}, where one needs for
converged NN observables at least $j_{max}$~=~14 at 1~GeV, we estimate
that a converged partial wave 3N calculation of the three-boson system
would need $J_{max}=l_{max}+\lambda_{max}$~=~14+15~=~29.

Let us now regard the two cases, (a) $E_{lab}$~=~200~MeV,
$l_{max}$~=~5, $\lambda_{max}$~=~7, $J_{max}$~=~12, and (b)
$E_{lab}$~=~1~GeV, $l_{max}$~=~14, $\lambda_{max}$~=~15,
$J_{max}$~=~29 in a partial wave decomposition. To illustrate the
tremendous number of partial waves needed in (b) compared to the
feasible case (a) it is sufficient to consider simple algebra for
different values of $J$.  Take for example $J$~=~5. Then simple
counting yields 30 different $l-\lambda$ combinations in case (a) and
125 in case (b).  For $J$~=10 this number increases in case (b) to
160. Moreover, since the number of total $J$'s at 1~GeV is more than
twice the number of $J$'s at 200~MeV, it appears quite unreasonable to
enforce a partial wave decomposition at energies far above 200~MeV in
the 3-boson (nucleon) system.  In addition, it would also be
numerically very demanding to evaluate the various ingredients in the
Faddeev equation reliably for the very high angular momenta. Another
advantage of using direct integration of vector variables in the Faddeev
equation is the simplicity of the permuation operators.


\section{Results and Discussion}

In the following we present our results for elastic and breakup
scattering in the energy regime from about 200 to 1500~MeV laboratory
projectile kinetic energy. We start with a comparison of our model
calculation to calculations based on a realistic NN force at lower
energies in order to show that even though our model is very simple,
we see similar features in the cross sections. Then we study
relativistic effects at higher energies. There are several questions
we want to address.  First,  we want to identify
scattering observables that are sensitive to the difference between
the relativistic and non-relativistic formulations of the three-body
problems and
to study the size of those relativistic effects as function of
increasing energy. This can at present only be done with our model
interaction. Second, we want to study the convergence properties of
the Faddeev multiple scattering series as function of the projectile
kinetic energy. Here the question of interest is, if, once the energy is
high enough, it is sufficient to only consider the first few terms in
the multiple scattering series. In addition, we also want to study
some approximations to our relativistic scheme.

\subsection{Comparison with calculations based on a realistic NN 
interaction at 200~MeV}

The laboratory kinetic energy of 200~MeV is a perfect energy to study
if the features of the 3N system we find based on our model
interaction are also present in calculations based on a realistic
model of the NN interaction, which describes the NN observables with
high accuracy.  The so-called high-precision interactions are fitted
up to about 350~MeV, but strictly speaking only valid below the
pion-production threshold.  We also know \cite{Witala1,Witala2,Witala3} that
relativistic effects are already visible at 200~MeV.

\begin{figure}
\begin{center}
 \includegraphics[width=7.1cm]{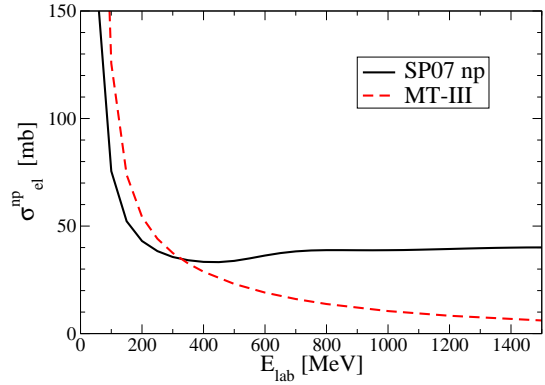}
\end{center}
\caption{
(Color online) The neutron-proton differential cross section as a function of the projectile
laboratory kinetic energy. The solid line represents the `experimental' cross
section obtained from the SAID data base \protect\cite{SAID}, and the dashed
line shows the two-body cross section obtained from the Malfliet-Tjon-III
potential \protect\cite{t-matrix} used in our calculations.
\label{fig3}}
\end{figure}

We choose the CD-Bonn interaction~\cite{CDBONN} for this comparison.
In Fig.~\ref{fig3} we show the $np$ total cross section extracted from
the SAID database~\cite{SAID} together with the total cross section
obtained from the MT-III interaction assuming bosonic symmetry. The
parameters of the MT-III interaction~\cite{Hang1} are adjusted such
that a two-body bound state at $E_d$~=~2.23~MeV is
supported. Fig.~\ref{fig3} shows that the experimental $np$ total
cross section falls slightly below the two-body cross section
predicted by our model at energies smaller than 
 $E_{lab} \simeq$~300~MeV, is about equal
between 300 and 400~MeV, and then reaches a constant value from about
600~MeV on, while our model prediction continues to decrease. The
slight rise of the experimental value around 600~MeV is a
manifestation of the influence of the $\Delta$(1232) resonance in the
NN system. The CD-Bonn interaction is fitted to NN observables to
about 350~MeV laboratory projectile energy and thus coincides with the
SAID result up to that energy.

\begin{figure}
\begin{center}
 \includegraphics[width=15.0cm]{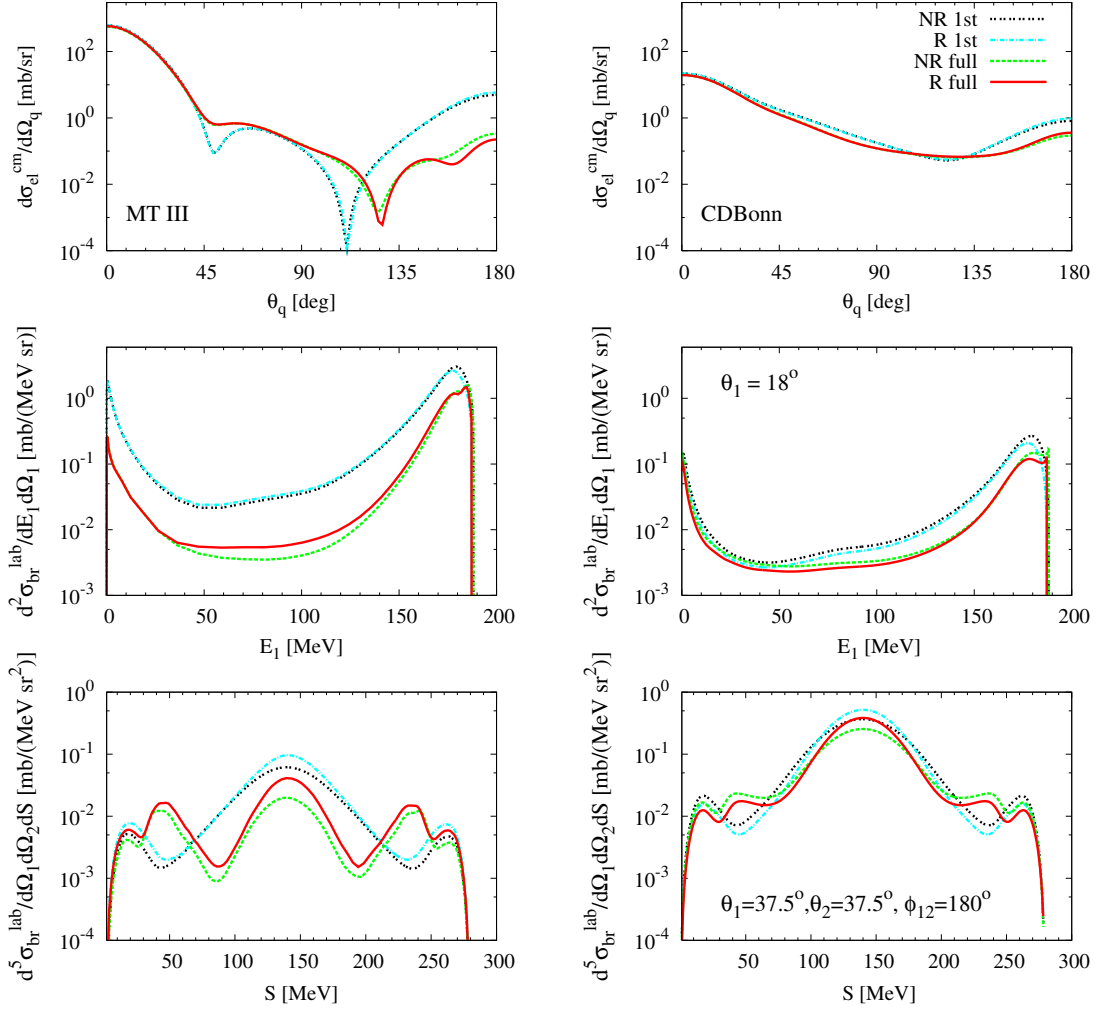}
\end{center}
\caption{(Color online) Three nucleon scattering at laboratory projectile kinetic energy
$E_{lab}$~=~200~MeV. The left column shows results obtained from the
Malfliet-Tjon-III potential assuming boson symmetry
and no partial wave decomposition, the right column shows
the corresponding realistic calculations obtained with the CD-Bonn
\protect\cite{CDBONN} potential 
where partial wave decomposition is applied. 
The top row displays the differential cross
section for elastic scattering, the middle row shows the breakup cross
section for inclusive scattering for the laboratory angle $\theta_1 = 18^o$
of the outgoing particle. The bottom row shows
the five-fold differential for exclusive breakup reaction as a function of
the arc-length S. The laboratory angles of the outgoing particles are
$\theta_1 = \theta_2 = 37^o$, and $\phi_{12}= 180^o$. The fully relativistic
converged Faddeev calculations are given by the solid lines (R), the
corresponding nonrelativistic calculations by the long-dashed lines (NR). 
In addition the relativistic (dash-dot) and nonrelativistic
(dotted) first order calculations are shown.
\label{fig4}}
\end{figure}

In Fig.~\ref{fig4} we show a comparison of elastic and breakup cross
sections at 200~MeV projectile laboratory energy for the 3D calculations
based on our MT-III model interaction 
and calculations based on a partial wave decomposition
employing the CD-Bonn potential.
The top row displays the differential cross section for elastic
scattering. We see that in both
cases the difference between the fully relativistic calculation and
the nonrelativistic one is overall quite small, and mostly visible at
the backward angles, an observation already made in
\cite{Witala1}. The differential cross section in the
forward direction is much larger for our model interaction, which is
consistent with the larger two-body total cross section.  In addition,
there are more diffraction minima in the bosonic case than in the
fermionic case, however, the minimum at around 130$^o$ is present in
both calculations. In the middle row we display the cross section for
inclusive breakup scattering as function of the laboratory kinetic
energy of the ejected particle at fixed laboratory angle 18$^o$. Both
cross sections are qualitatively similar, the fully converged Faddeev
calculation gives a lower cross section than the first order
calculation, indicating the importance of rescattering contributions
at this low energy. The difference between the relativistic and
nonrelativistic calculations is quite small in both cases. In the
calculation based on the CD-Bonn interaction the FSI peak is more
pronounced due to the virtual bound state in the $^1$S$_0$ state. The
latter is absent in the MT-III model. The bottom row shows the
five-fold differential cross section as function of the arc-length $S$
for a configuration in which the laboratory angles $\theta_1 = \theta_2
= 37^o$ are given in the scattering plane ($\phi_{12}=180^o$). The
position of the peaks is identical for both calculations, which is a
manifestation that peak structures are given by the kinematics of the
problem.  
In both cases the relativistic calculation gives a significantly 
larger
cross section for the central peak at $S \approx 140$~MeV compared to the
nonrelativistic result, an 
increase by a factor of $\sim$1.5 for the full partial wave  
calculation and by a factor $\sim$ 2 for the full 3D calculation.
This increase is already present in both first order calculations.
For the MT-III model this trend is the
same for all other peaks, whereas for the CD-Bonn model the
nonrelativistic calculations give a slightly larger cross section
compared to the relativistic one in the peaks at small and large
 values of arc-length S.

Summarizing, the comparison of cross section obtained from our model
interaction MT-III with those given by a realistic NN interaction like
CD-Bonn at 200~MeV indicates that, despite our model being quite
simple, the qualitative features of especially the breakup cross
sections are very similar. The differences between the fully
relativistic calculations and their nonrelativistic counterparts are
still quite small at this low energy  for elastic scattering and 
inclusive breakup. For the exclusive breakup, 
however, even at this energy complete configurations with large changes 
of the nonrelativistic cross section due to relativity can be found. This 
sensitivity of the complete breakup to relativistic effects has already been 
observed in~\cite{Witala2,Witala3}.


\subsection{Elastic Scattering at Intermediate Energies}

\begin{figure}
\begin{center}
 \includegraphics[width=9.0cm]{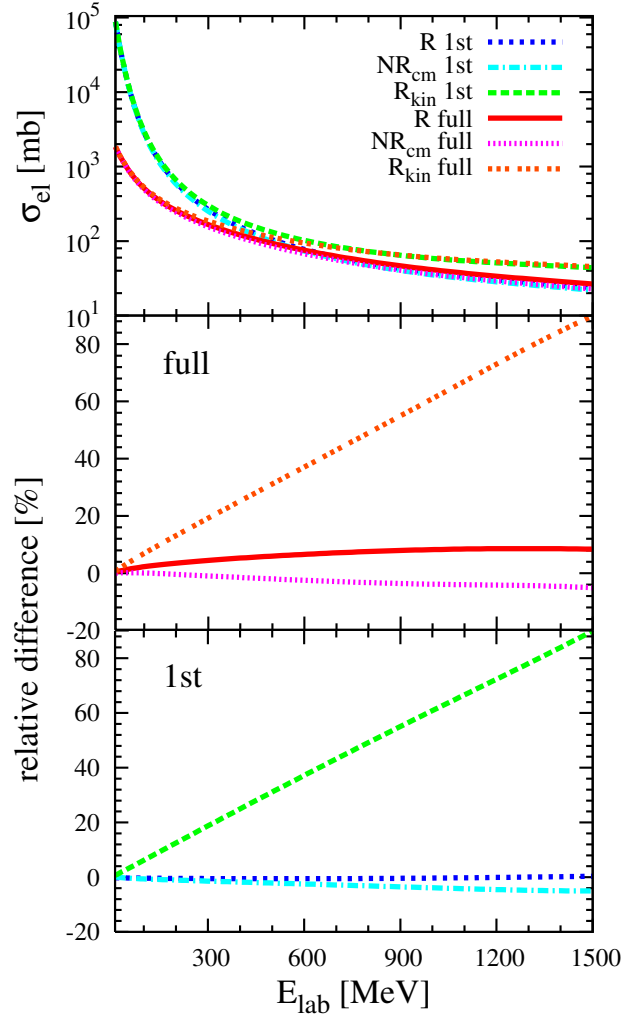}
\end{center}
\caption{(Color online) The total cross section for elastic scattering as a 
function of the projectile
kinetic energy (top panel). The fully relativistic Faddeev calculation is shown as
solid line, the corresponding first order term by the short dashed line. Calculations
which only use relativistic kinematics, i.e. the Lorentz transformation between 
laboratory and c.m. frame together with the relativistic phase space factor (labeled
R$_{\rm kin}$) are
given as dotted line for a full Faddeev calculation and as short-dashed line for the
first order term. Calculations which only take into account the Lorentz transformations
between the laboratory and c.m. frame (labeled NR$_{\rm cm}$) are shown as
dotted line for the full Faddeev calculation and as dash-dotted line for the 
1st order one. The middle panel shows the relative difference between the fully
relativistic Faddeev calculation and the nonrelativistic one (solid line) together
with the difference to the nonrelativistic calculation if only relativistic
kinematics is considered. The bottom panel shows the corresponding relative
differences when only the first order term is taken into account.
\label{fig5}}
\end{figure}

Starting from our model interaction we now consider three-body
scattering in the energy regime up to 1.5~GeV. The total cross section
for elastic scattering is related to the symmetrized 
transition operator $U$ of
Eq.~(\ref{eq:2.1})  via
\begin{equation}
\sigma_{el}= (2\pi)^4 \int d\Omega \frac{E_n^2 (q_0) E_d^2(q_0)}
{{\sf W}^2}
|\langle \varphi_d, {\bf {\hat q}} q_0|U|\varphi_d, {\bf q_0}\rangle|^2.
\label{eq:3.1}
\end{equation}
In Fig.~\ref{fig5} we display the total cross section for elastic
scattering as a function of the projectile kinetic energy up to 1.5~GeV
obtained from our fully converged relativistic Faddeev calculation as
well as the one obtained from the first order term. It is obvious 
that,
especially for the energies below 300~MeV, the contribution of
rescattering terms is huge.  Since the logarithmic scale from the top
panel is unsuited to extract detailed information about the size of
relativistic effects, we show in the lower two panels the relative
difference of the relativistic calculations with respect to their
nonrelativistic counterparts.  The bottom panel displays the relative
difference between the relativistic first order term and its
nonrelativistic counterpart as dotted line. Essentially the first order
calculation does not show any effect. This is theoretically consistent
when having in mind that in first order ($T^{1st}=tP$) only the
two-body t-matrix enters into the cross section. The relativistic
two-body t-matrix is constructed to be phase-shift equivalent to the
non-relativistic one via the CPS method
\cite{cps,Keister:2005eq}. Thus, seeing no difference between the
fully relativistic and the corresponding nonrelativistic calculations
indicates that relativistic effects are taken into account
consistently at the two-body level. Doing the same comparison with
fully converged Faddeev calculations (solid line in the middle panel)
indicates that relativistic effects in the three-body problem increase
the elastic scattering 
 total cross section with increasing energy. At our highest energy,
1.5~GeV, this increase is about 8.3\%.

Often only effects due to relativistic kinematics are taken into
account. Here we have the opportunity to study consequences of such a
simple approximation. For the calculations labeled $R_{kin}$ we only
consider the Lorentz transformations from laboratory to
center-of-momentum (c.m.) frame and the relativistic
phase space factor of Eq.~(\ref{eq:3.1}), whereas the matrix elements
of the operator $U$ are calculated from the solution of the
nonrelativistic Faddeev equation. The relative difference between this
calculation and a completely nonrelativistic calculation is indicated
by the short-dashed line in the bottom panel of Fig.~\ref{fig5}, where
only the first order term is considered. The triple-dotted curve in the
middle panel is the same comparison, but now between fully converged
Faddeev calculations.  For both, full and first order calculation, the
effect is huge. To understand better which piece of the kinematics
included is responsible for this large enhancement of the cross
section, we also plot in Fig.~\ref{fig5} calculations (labeled
NR$_{\rm cm}$), which only contain the Lorentz transformation between
the laboratory and c.m. frame, but carry the nonrelativistic phase
space factor in Eq.~(\ref{eq:3.1}). The dash-dotted line in the lower
panel show the 1st order calculation and the dotted line in the middle
panel the full Faddeev calculation.  Using a Lorentz transformation in
the change of frames has the effect that the two-body t-matrix is
calculated at a slightly different c.m. momentum ${\bf q_0}$, and thus
there is a small effect, about a 5\% underestimation of the total
cross section.  The huge effect is entirely due to the relativistic
phase space factor, and relativistic dynamics then has an equally
large effect of the opposite sign. 
This interplay of increasing effects due to the relativistic phase-space 
factor and decreasing effects due to relativistic dynamics has been already
observed in the partial-wave based Faddeev calculations with 
realistic interactions \cite{Witala1}. It led for 
elastic scattering cross section at 250~MeV to relativistic 
effects which are relatively 
 small and restricted to backward angles.
Recent measurements of the neutron-deuteron ($nd$) differential cross section
 at 248~MeV~\cite{Maeda:2007zza}
indicate that for discrepancies of theoretical predictions
 in this observable,  
short range components of a three-nucleon force are
equally important. 

The problem with approximating relativistic effects only through
kinematics and phase space factors can be easily
understood in the 2+1 body problem, where the phase equivalence is
achieved by choosing the invariant mass as a function of the
non-relativistic two-body Hamiltonian, $M = f (h)$.  The eigenvalues
equation for the scattering problems
\begin{equation}
\vert \psi \rangle = {1 \over f(w) -f(h_0)+i0^+ }(f(h)-f(h_0))
\vert \psi \rangle \qquad     
\vert \psi \rangle = {1 \over w -h_0 +i0^+ } (h -h_0)
\vert \psi \rangle \qquad     
\end{equation}
are equivalent, but the replacement of $f(w)$ by $w$ must be compensated by 
replacing the interaction $f(h) - f(h_0)$ by $h-h_0$.  Including only
kinematic relativistic effects is equivalent to making the replacement
$h_0 \to f(h_0)$ without making the compensating replacement 
$v= h-h_0 \to f(h) - f(h_0)$.

\begin{figure}
\begin{center}
 \includegraphics[width=15.0cm]{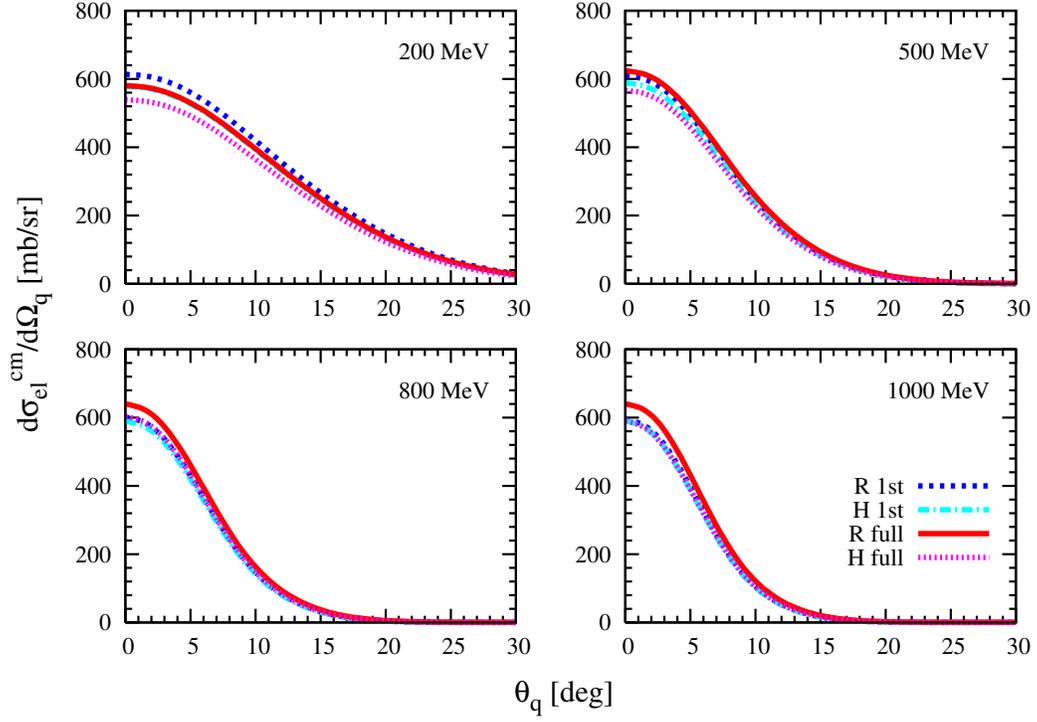}
\end{center}
\caption{(Color online) The differential cross section for elastic scattering as a
function of c.m. angle
$\theta_q$  for selected laboratory kinetic energies. 
The converged solution of the relativistic Faddeev equation is given as solid line. The
dotted line shows the converged solution of the relativistic Faddeev equation in which
the fully off-shell 2N t-matrix is replaced by the half-shell t-matrix. 
The corresponding first order calculation are given by the short-dashed line and the
dash-dotted line. For details see text.  
\label{fig6}}
\end{figure}

We also study a more sophisticated approximation to the relativistic
dynamics. In Ref.~\cite{Lin:2007ck} we described in detail how we
obtain the transition amplitude of the 2N subsystem, $T_1({\bf k},{\bf
k'},{\bf q};z)\equiv \langle {\bf k}|T_1({\bf q};z)|{\bf
k'}\rangle$, embedded in the three-particle Hilbert space which
enters the Faddeev equation, Eq.~(\ref{eq:2.10}).  The fully off-shell
amplitude is the solution of a first resolvent type 
equation~\cite{Keister:2005eq} given by
\begin{eqnarray}
\langle \mathbf{k} |T_1(\mathbf{q};z)| \mathbf{k}' \rangle  &=&
\langle \mathbf{k} | T_1(\mathbf{q};z')| \mathbf{k}' \rangle \; - \nonumber \\
\int & d\mathbf{k}''& \langle \mathbf{k} | T_1(\mathbf{q};z)|
\mathbf{k}'' \rangle   \Big( \frac{1}{z - \sqrt{4(m^2 + \mathbf{k}''^2)
+ \mathbf{q}^2} }  - \frac{1}{z' - \sqrt{4(m^2 +
\mathbf{k}''^2) + \mathbf{q}^2} } \Big) \langle \mathbf{k}'' |
T_1(\mathbf{q};z') | \mathbf{k}' \rangle.
\label{eq:3.2}
\end{eqnarray} 
Here $T_1(z')$ is taken to be right half-shell with $z'=\sqrt{4(m^2 +
\mathbf{k}'^2) + \mathbf{q}^2} +i\epsilon$.  Note that in this
equation the unknown matrix element is to the {\it left} of the
kernel.  It was suggested in Ref.~\cite{Keister:2005eq} that a
reasonable approximation to this embedded 2N transition amplitude
might by the Born term of the above integral equation, 
which is
\begin{equation}
\langle \mathbf{k} |T^H_1(\mathbf{q};z)| \mathbf{k}' \rangle \simeq  \langle \mathbf{k} |
T_1(\mathbf{q};z')| \mathbf{k}' \rangle .
\label{eq:3.3}
\end{equation}
In this approximation, the fully off-shell 2N transition amplitude is
replaced by a half-shell amplitude.  The effect of this approximation
is not large in elastic scattering, as shown in Fig.~\ref{fig6}, where
we plot the differential cross section in forward direction for the
fully relativistic calculations and the ones containing the
approximation of Eq.~(\ref{eq:3.3}) to the boost (curves labeled
H). Consistently and independent of projectile energy, approximating
the embedded two-body t-matrix by  the half-shell t-matrix leads to
an underprediction of the differential cross section in forward
direction. Though not plotted, this also leads to a smaller total cross
section for elastic scattering. 

Finally, we want to investigate the convergence of the multiple
scattering series as a function of the projectile laboratory kinetic
energy. One might expect that with increasing energy only a few terms
in the multiple scattering series are sufficient for a converged
result. Our converged relativistic Faddeev calculations now allow a
detailed study. This is of particular interest, since recently
relativistic calculations in the energy regime around 1~GeV have been
published \cite{Ladygina:2007ud,Ladygina:2004rr,Ladygina:2003kj},
which are carried out in a multiple scattering expansion of the
Faddeev equations up to 2nd order, and which use the off-shell
continuation of the experimental NN amplitudes as two-body input.

\begin{figure}
\begin{center}
\includegraphics[width=15.5cm]{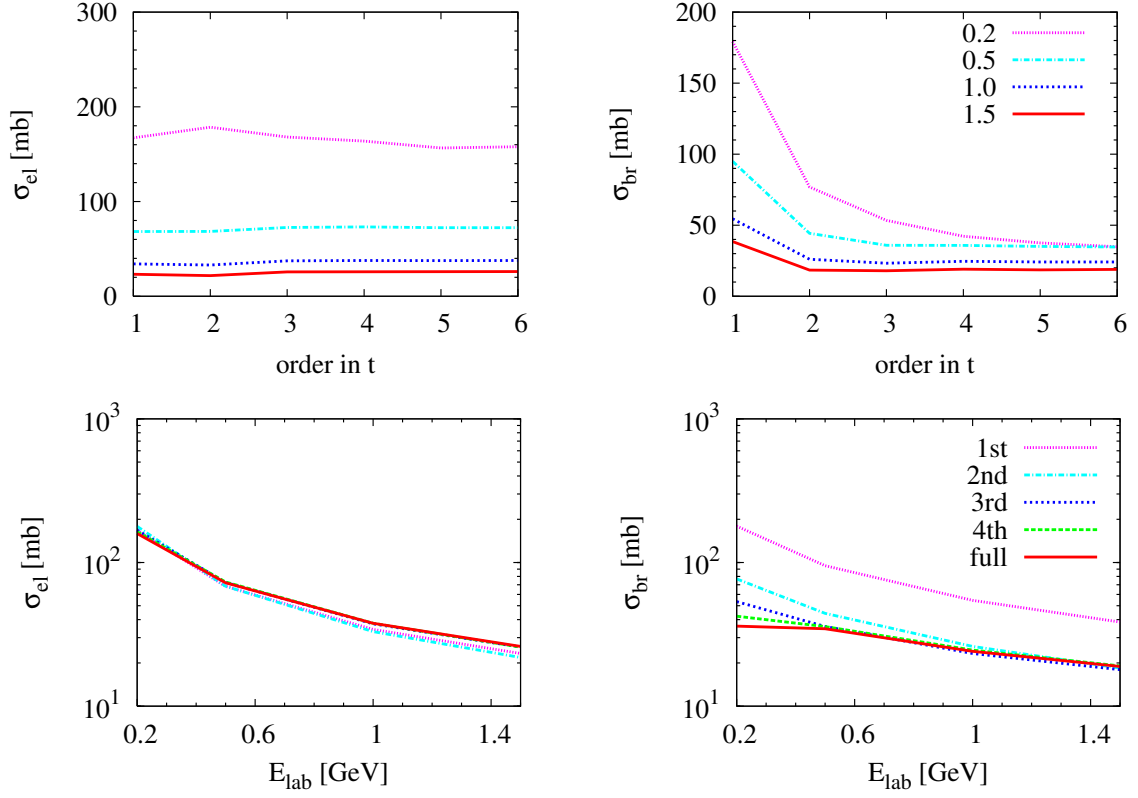}
\end{center}
\caption{(Color online) The total cross section for elastic scattering (left column) and for breakup
reaction (right column). The bottom row shows both relativistic
 cross sections as function of
the projectile laboratory kinetic energy, when starting from the first order in the
Faddeev calculation successively the next 3 orders are added, together with the fully 
converged calculation (solid line).  The top row displays the change in both total
cross sections as a function of the order in the multiple scattering series for
selected laboratory projectile energies (given in the legends of the top right
panel in units of GeV).
\label{fig7}}
\end{figure}

First we want to consider the convergence of the Faddeev multiple
scattering series in the total cross sections for elastic scattering as well as
breakup reactions as a function of projectile kinetic energy. In the
bottom row of Fig.~\ref{fig7} the different orders (successively summed
up as Neumann sum to the order indicated in the legend) are shown as
functions of the projectile laboratory energy. We see a distinct
difference in the behavior of the elastic total cross section in
comparison with the breakup total cross section. While the elastic
total cross section converges very rapidly, the total breakup cross
section does not.  The left upper panel of Fig.~\ref{fig7} shows the
elastic total cross section as a function of the order in the multiple
scattering series (the orders are successively summed up the one
indicated on the x-axis). Even at 200~MeV there is very little change
due to contributions from the 2nd or higher order rescattering
terms.  For the higher energies, the 1st order term already captures
the essential physics. This is very different for the total breakup
cross section, where for 200~MeV projectile energy the full solution
of the Faddeev equation is clearly necessary. For energies of 1~GeV
and higher, at least one rescattering contribution (2nd order in the
multiple scattering series) is necessary to come close to the full
solution.

\begin{figure}
\begin{center}
\includegraphics[width=9.0cm]{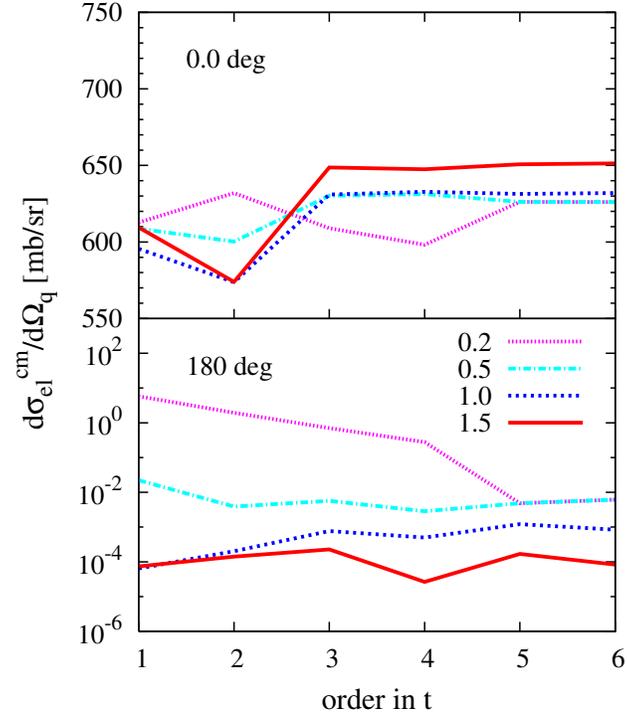}
\end{center}
\caption{(Color online) The differential cross section for elastic scattering in forward
(0 deg) and backward (180 deg) direction as a function of the
order in the multiple scattering series for selected projectile laboratory 
kinetic energies indicated in the legend in units of GeV. 
\label{fig8}}
\end{figure}

Since the total cross section for elastic scattering might be
insensitive to higher orders in the Faddeev multiple scattering
series, we plot in Fig.~\ref{fig8} the differential cross section at
forward and backward angles as a function of the order in the multiple
scattering series for the same laboratory projectile energies. Here we
see that at the lowest energy, 0.2~GeV, the convergence is not as fast
as the total cross section suggests. In fact, at least 5 orders are
necessary, which is consistent with the experience from
nonrelativistic calculations at low energies \cite{wgphysrep}.  For
energies of 1~GeV and higher, the forward direction is converged at
the 3rd order in $t$, whereas the backward angle is not as sensitive
(it should be pointed out that the cross section in backward direction
is about five orders of magnitude smaller than the one in forward
direction). It seems accidental that the multiple scattering series
converges faster at 0.5~GeV compared to 1~GeV. However, a similar
finding was presented in Ref.~\cite{Ladygina:2007ud}, where it was
observed that polarization observables for elastic proton-deuteron
($pd$) scattering at 395~MeV were described better than those at
1.2~GeV, when calculating the Faddeev multiple scattering series up to
the 2nd order.


\subsection{Breakup Scattering at Intermediate Energies}

The calculation of breakup cross sections requires knowledge of
the matrix element $\langle {\bf k},{\bf q} \vert U_0\vert \varphi_d,
{\bf q_0}\rangle$ in Eq.~(\ref{eq:2.1}). For details of the derivation
we refer to Ref.~\cite{Lin:2007ck} and only give the final expressions
here. The five-fold differential cross section for exclusive breakup
is given in the laboratory frame as \cite{Lin:2007kg}
\begin{eqnarray}
\frac {d^5\sigma_{br}^{lab}}{ d\Omega _1d\Omega _2dE_1} 
&=& (2\pi)^4 \frac{E(q_0) E_d(q_0)  E(q)}{2k_{lab}m_d } 
\frac{p_1p_2^2} {p_2({\sf E}-E(p_1)) - E(p_2)
(\mathbf{P}-\mathbf{p}_1) \cdot \hat {\mathbf{p}}_2 } 
\nonumber \\
& & \times E(k) \sqrt{4 E^2(k)+\mathbf{q}^2} \;
\left| \langle \mathbf{k}, \mathbf{q}  \Vert U_0 \Vert
\varphi_d, \mathbf{q}_0 \rangle  \right| ^2 .
\label{eq:3.4}
\end{eqnarray}
Here ${\sf E}$ is the total energy of the system and $\mathbf{P}$ its
total momentum. The subscripts $1$ and $2$ indicate the two outgoing
particles.  In inclusive breakup only one of the particles is
detected, and thus one of the angles in Eq.~(\ref{eq:3.4}) is
integrated out.  This leads to the inclusive breakup cross section in
the laboratory frame \cite{breakupcorr}
\begin{eqnarray}
 \frac{d^3 \sigma_{br}^{lab}}{d\Omega _1 dE_{1}} 
&=& (2\pi)^4  \frac{{\sf E}}{{\sf W}} \frac {E(q_0) E_d(q_0)}{4k_{lab}m_d}
 \frac{p_1k_aE(q)(4 E^2(k_a)+\mathbf{q}^2) }{\sqrt{4 E^2(k_a)+(\mathbf{P} - \mathbf{p}_1)^2}} 
 \int d\Omega _k | \langle \mathbf{k}_a \mathbf{q}  | U_0 | \varphi_d \mathbf{q}_0 \rangle |^2 ,
\label{eq:3.5}
\end{eqnarray}
where $k_a$ is determined by the condition ${\sf
W}=\sqrt{4(m^2+\mathbf{k}_a^2)+\mathbf{q}^2}+\sqrt{m^2+\mathbf{q}^2}$.

In Refs.~\cite{Lin:2007ck,Elster:2007vg,Lin:2007kg} we already pointed
out and demonstrated that relativistic kinematics is essential to
obtain the correct position of e.g. the peak for quasi-free scattering
(QFS), especially at higher energies. The difference
between a nonrelativistic calculation of the breakup cross section and
a relativistic one is quite large at higher energies.  However one may
argue that this difference is artificially large, since it is natural
to use relativistic kinematics at higher energies. Therefore, here we
will {\it not} compare to entirely nonrelativistic calculations, but
rather calculations where the three-body transition amplitude has
been obtained from the solution of a nonrelativistic Faddeev equation,
but the transformations between the laboratory frame and the
c.m. frame are Lorentz transformations.  This is equivalent 
to comparing the relativistic and non-relativistic calculations in the
center-of-momentum frame. 

\begin{figure}
\begin{center}
\includegraphics[width=15.0cm]{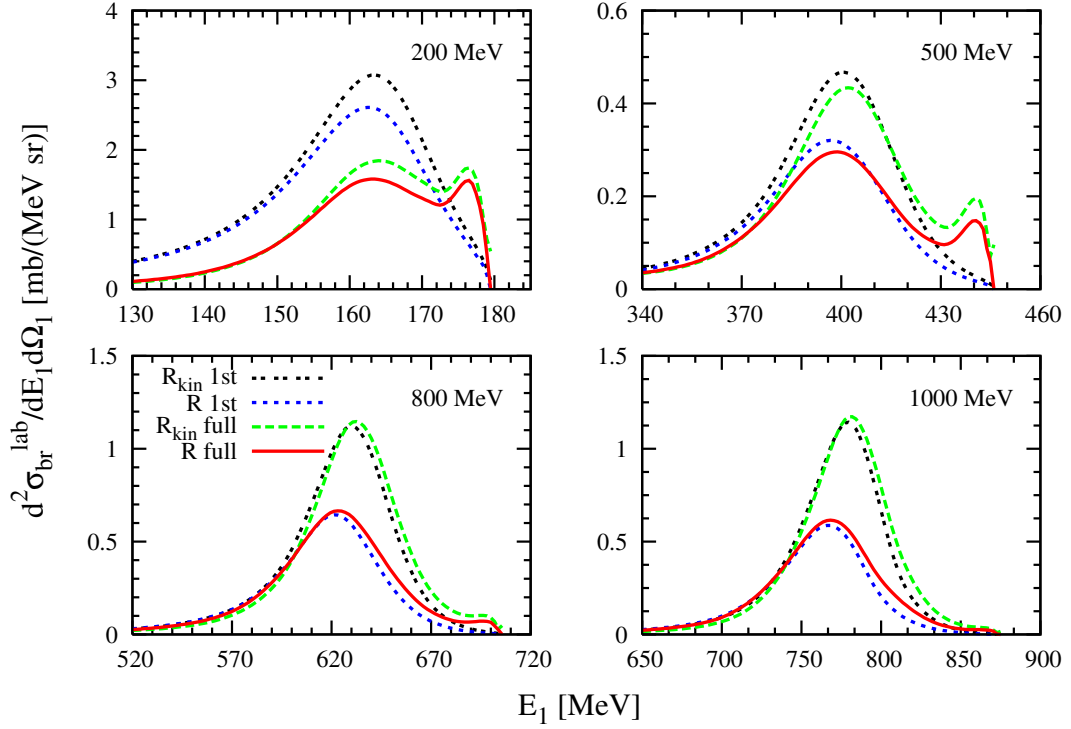}
\end{center}
\caption{(Color online) The inclusive breakup cross section as a function of the laboratory
 kinetic energy $E_1$ of the emitted particle at an emission angle $\theta_1 = 24^o$.
The incident laboratory kinetic energy for each cross section is indicated in 
each panel. The solid lines (R full) 
represent the converged relativistic Faddeev calculation
and the dotted line the corresponding first order calculations (R 1st). The lines
labeled R$_{\rm kin}$ correspond to calculations in which only relativistic
kinematics is taken into account.  
\label{fig9}}
\end{figure}

In addition we use the
relativistic phase space factor for the cross sections. In
Fig.~\ref{fig9} we show the inclusive breakup cross section as
a function of the laboratory kinetic energy of the ejected particle at
fixed angle $\theta_1 = 24^o$ for different projectile kinetic
energies calculated from the full solution of the relativistic Faddeev
equations together with `nonrelativistic' calculations using the above
defined relativistic kinematics.  There is still a shift of the
position of the QFS peak towards lower ejectile energies, which
increases with increasing projectile energy.  There is also a very
visible effect of the relativistic phase space factor used together with 
the nonrelativistic three-body transition amplitude. At 1000~MeV the
size of the QFS peak is a factor of two larger compared to exact
relativistic calculation.  For the lower energies the 1st order
calculation yields a significantly higher QFS peak compared to the
full calculation, whereas for the higher energies, the peak height is
almost the same for the 1st order and the full calculation.

\begin{figure}
\begin{center}
\includegraphics[width=15.0cm]{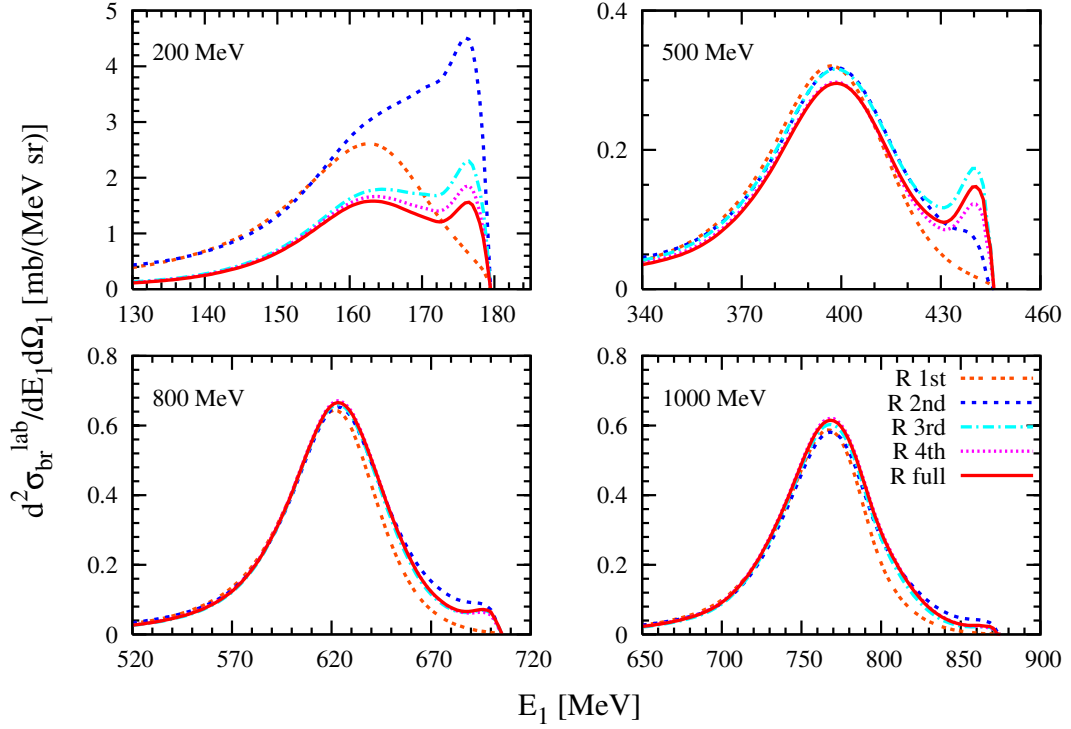}
\end{center}
\caption{(Color online) The inclusive breakup cross section as a function of the laboratory 
 kinetic energy $E_1$ of the emitted particle at an emission angle $\theta_1 = 24^o$.
The incident laboratory kinetic energy for each cross section is indicated in
each panel. The solid lines (R) represent the converged relativistic Faddeev
calculation. The triple-dotted line shows the 1st order calculation, for the short dashed
line the 2nd order contribution is added to the previous, for the dash-dotted line
the 3rd order is added, and for the dotted line the 4th order.
\label{fig10}}
\end{figure}

Next we investigate in detail the convergence of the Faddeev multiple
scattering series in the region of the QFS peak as a
function of the projectile energy. In Fig.~\ref{fig10} we display
calculations at selected energies from 200 to 1000~MeV. The solid line
represents the solution of the relativistic Faddeev equation, whereas
the other curves show the Neumann sum of the multiple scattering
series containing the sum up to the order in the two-body
t-matrix as indicated in the legend. For the lowest energy, 200~MeV,
it is obvious that the multiple scattering series does not converge
rapidly.  This changes considerably as the projectile kinetic energy
grows. Though the variation of the different orders is not as large
anymore at 500~MeV, the multiple scattering series must still be
summed up to 4th order in the QFS peak to coincide with the full
result, whereas at 800~MeV  already the 2nd order is almost identical
with the full result, and even a 1st order calculation can be
considered quite good. This trend continues as the energy grows. Of
course, 1st order calculations are never able to capture the FSI peak
at the maximum energy of the ejectile, nor do they describe the high
energy shoulder of the QFS peak. However, our study indicates that for
energies in the GeV regime it is very likely sufficient to consider only one
rescattering term when studying inclusive breakup 
 reactions in the vicinity of the QFS peak.

\begin{figure}
\begin{center}
\includegraphics[width=15.0cm]{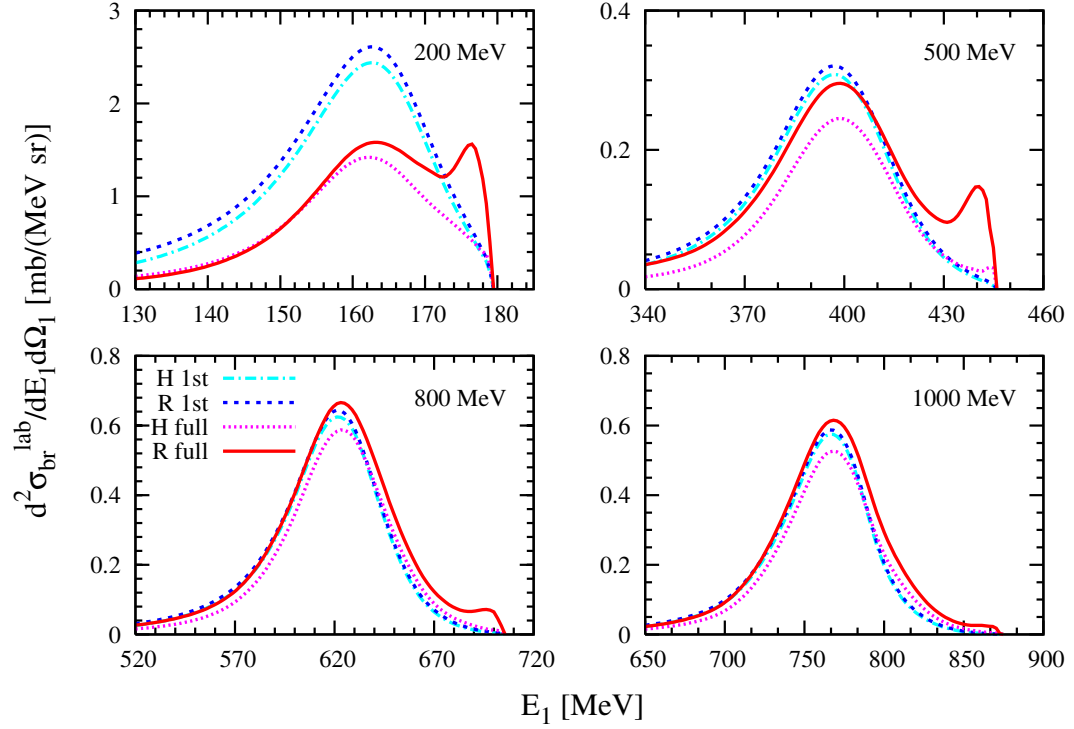}
\end{center}
\caption{(Color online) The inclusive breakup cross section as a function of the laboratory
 kinetic energy $E_1$ of the emitted particle at an emission angle $\theta_1 = 24^o$.
The incident laboratory kinetic energy for each cross section is indicated in
each panel. The solid lines (R) represent the converged relativistic Faddeev
calculation. The dotted line (H) displays the calculation in which the fully
off-shell two-body t-matrix is replaced by the half-shell one. The calculations
labeled 1st stand for the corresponding 1st order calculations.
\label{fig11}}
\end{figure}

Finally, we also want to study the approximation suggested in
Eq.~(\ref{eq:3.3}), namely replacing the off-shell two-body transition
amplitude embedded in the three-body Hilbert space by the half-shell
one. The calculations based on the approximation of Eq.~(\ref{eq:3.3})
and labeled (H) are plotted in Fig.~\ref{fig11} together with the
exact solution. Considering only the 1st order calculation we observe
a similar trend as in the differential cross section for elastic
scattering, the approximation slightly underpredicts the exact result,
independent of the energy under consideration. However, when this
approximate two-body transition amplitude is iterated to all orders in
the Faddeev equation, the deviations from the exact calculations
become larger. At 800 and 1000~MeV the iteration of the exact
amplitude increases the cross section in the QFS peak, whereas it
decreases for the approximation with respect to the 1st order term. At
200 and 500~MeV the approximation does not only give a smaller cross
section in the QFS peak but also fails to develop an FSI peak towards
the maximum allowed ejectile energies.  From this we conclude that
Eq.~(\ref{eq:3.3}) does not provide a good approximation for inclusive
breakup cross sections. Our calculations indicate that at energies
1~GeV or higher, it is important to carry out the Poincar\'e invariant
aspects of the calculation exactly. They also indicate that it is
sufficient to  consider only one rescattering term to capture most
features of the cross section.  Although these conclusions are based
on the use of a simple model two-body interaction, we conjecture that
calculations based on realistic interactions will have similar
characteristics.

\begin{figure}
\begin{center}
\includegraphics[width=15.0cm]{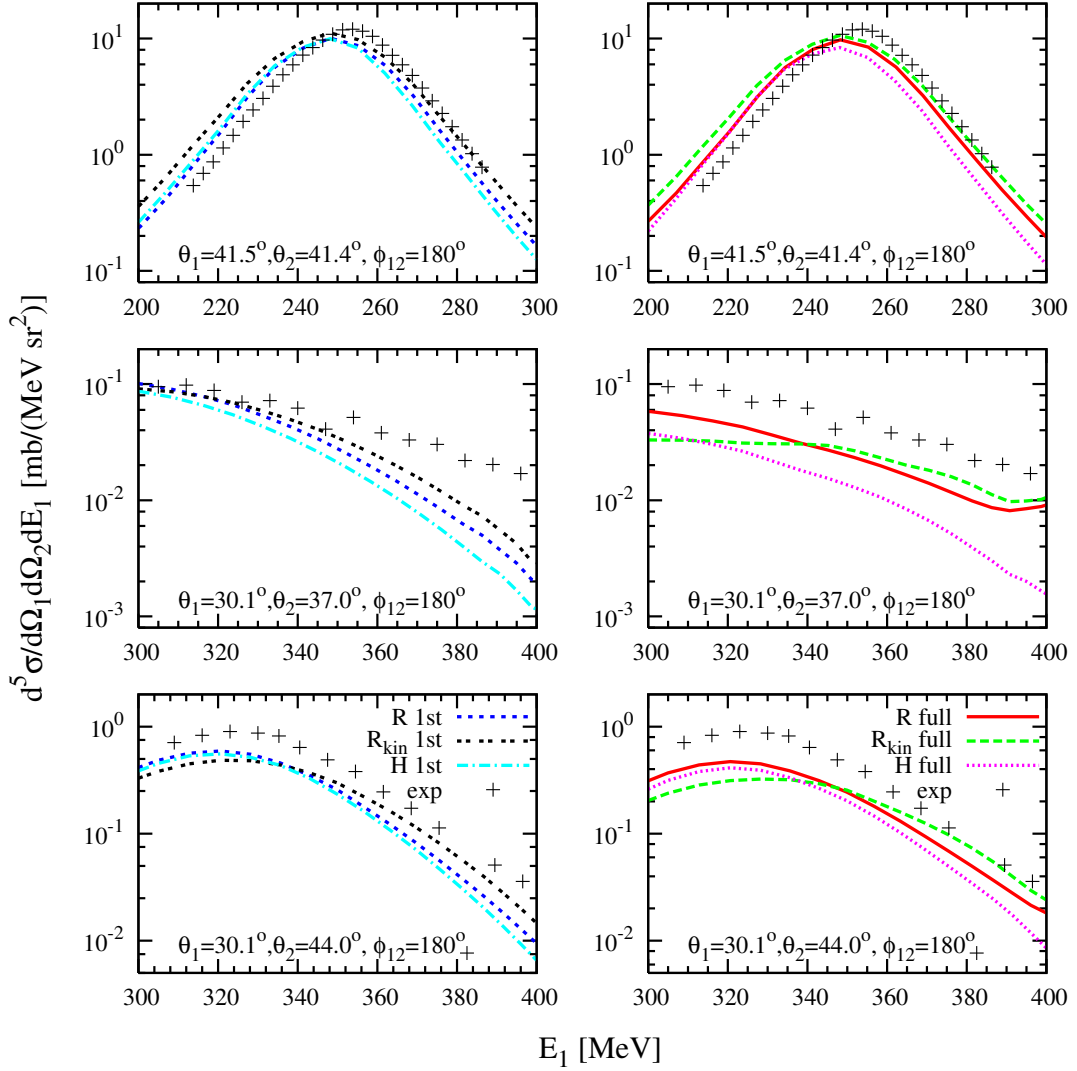}
\end{center}
\caption{(Color online) 
The exclusive differential cross section for the $^2$H(p,2p)n
reaction at 508~MeV laboratory projectile energy  for different
proton angle pairs $\theta_1$-$\theta_2$ with respect to the beam axis
as a function of the laboratory kinetic energy of the first detected proton.
The left column represents 1st order calculation, whereas the right column gives
full solution of the Faddeev equation. The curves labeled R (solid for the full Faddeev
calculation and dotted for the 1st order one) represent the full relativistic
calculations, whereas for the curves labeled R$_{\rm kin}$ (dashed in the right column
and double-dotted in the left)  only relativistic kinematics
is taken into account (see text), and for the curves labeled H (dotted in the right
column and dash-dotted in the left) the fully off-shell t-matrix is replaced by the
half-shell one. The data are
taken from Ref.~\protect\cite{Punjabi:1988hn}.
\label{fig12}}
\end{figure}

For our study of exclusive breakup scattering in the intermediate
energy regime we choose two different experimental situations where
there are data available.  First we consider the $^2$H(p,2p)n reaction
at 508~MeV, where the two outgoing protons are measured for a given
angle pair $\theta_1 - \theta_2$ in the scattering
plane~\cite{Punjabi:1988hn}. Since the convergence of the multiple
scattering series is already discussed in \cite{Lin:2007kg}, we only
want to investigate the effect of the approximations previously given
in this reaction. In Fig.~\ref{fig12} selected angle configurations
are shown. The left column of the figure shows the first order
calculations and the right column the full solution of the Faddeev
equation.  The exact 1st order calculation is given by the dotted line
in the left column and the exact full solution by the solid line in
the right column. The angle combination $\theta_1 = 41.5^o - \theta_2
= 41.4^o$ is a QFS configuration. First, we see that in a QFS
configuration, the 1st order calculation is already almost identical
to the full Faddeev calculation~\cite{Lin:2007kg}, whereas this is not
the case for the other configurations shown. If only relativistic
kinematics is considered, namely the Lorentz transformations between
laboratory and c.m. frame together with the relativistic phase space
factor, and a nonrelativistic three-body transition amplitude is
employed, we obtain the double-dotted curve for the 1st order 
calculations and the dashed line for the full solution of the Faddeev
calculation.  Again, the QFS configuration is quite insensitive to
this approximation. However, the deviation from the exact calculation
is quite visible in the other two configurations shown. Finally, we
also consider the approximation suggested by Eq.~(\ref{eq:3.3}), which
is indicated by the dash-dotted line, labeled `H' in the left column
(1st order calculation) and the dotted line in the right column
(full solution of the Faddeev equation). Here we see that even in the
QFS configuration there are already deviations of this approximation
for the high energy shoulder. The approximation underpredicts the full
solution. This tendency becomes stronger for the other two
configurations.
The interesting property 
of this approximation is that while it appears to be a reasonable 
approximation to the Faddeev kernel, the errors in the approximation 
increase when the equation is iterated.
Thus we conclude that this approximation, though simplifying the
calculation of the two-body t-matrix embedded in the three-body
Hilbert space, does not seem to capture essential structures of the
two-body t-matrix. 
The failure of this approximation, which approximates the 
off shell two body transition operator in the Faddeev equation with the 
half-shell transition operator,  suggests that some care
is necessary in modeling the off-shell behavior of the transition 
operators in more phenomenological schemes.

\begin{figure}
\begin{center}
\includegraphics[width=15.0cm]{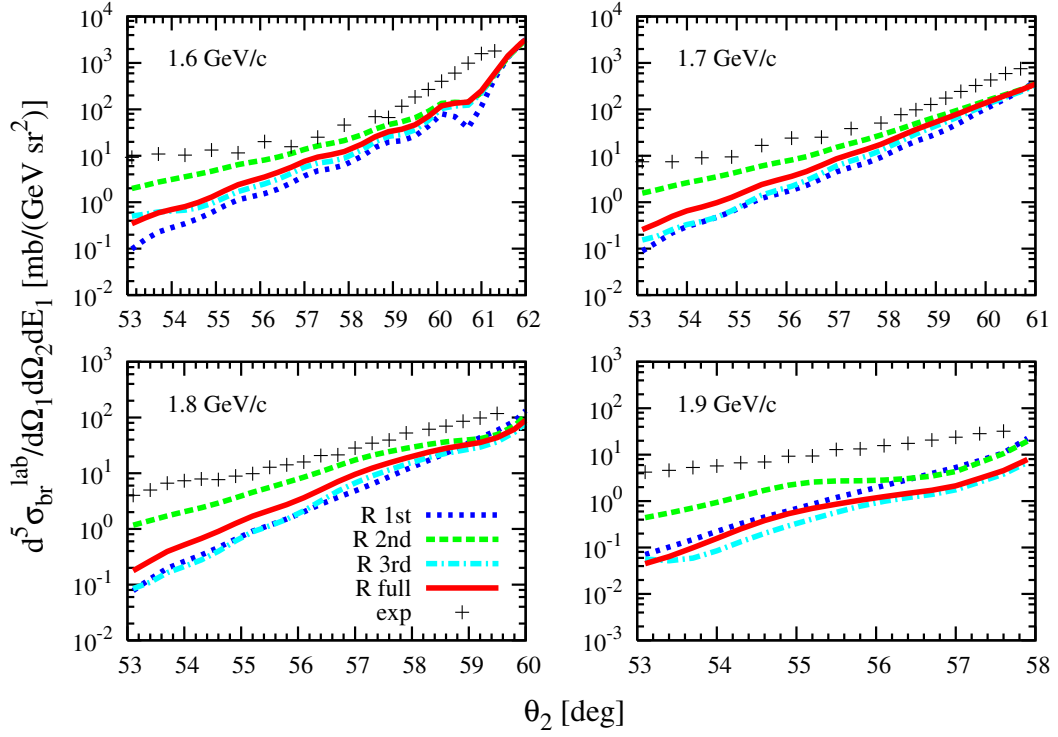}
\end{center}
\caption{(Color online) The exclusive differential cross section for the
reaction $^1$H(d,2p)n at 2~GeV deuteron energy as a function of the angle $\theta_2$ of
the second  of the outgoing protons for a fixed first proton momentum indicated 
in the figure. The
solid line represents the solution of the full relativistic Faddeev equation. 
The dotted line gives the result of the first order calculation. For the dashed 
line the 2nd order
term is added and  for the dash-dotted line the 2nd and 3rd order terms 
are added. 
The data are taken from Ref.~\protect\cite{Ero:1993aw}.
\label{fig13}}
\end{figure}

For breakup reaction at a slightly higher energy we
consider the $^1$H(d,2p)n reaction at 2~GeV deuteron kinetic energy
\cite{Ero:1993aw}. Here the two outgoing protons are measured.
Energetically, this reaction would correspond to $pd$ scattering at
roughly 1~GeV and thus is within the range of the calculations
presented here.  In Fig.~\ref{fig13} we show the five-fold
differential cross section as a function of the angle of  the second 
detected proton for four different  momenta of the first detected proton.  
The full
relativistic Faddeev calculation is represented by the solid line. In
order to investigate the convergence of the multiple scattering series
we show the 1st order calculation as a dotted line, then
successively add one (2nd order) and two (3rd order) rescattering
terms to the leading order. In this reaction, the first two
rescattering terms are about the same size, but have opposite sign, so
that the 3rd order calculations are very close to the 1st order
one. We also observe that the 3rd order calculation is already so
close to the full Faddeev calculation that the Neumann series can be
considered converged with three terms.


\section{Summary and Conclusion}

In this work we demonstrated the feasibility of applying Poincar\'e
invariant quantum mechanics to model three-nucleon reactions at
energies up to 2 GeV.  This is an important first step for studying
dynamical models of strongly interacting particles in the energy range
where sub-nuclear degrees of freedom are thought to be relevant.  At
these energies the Poincar\'e invariance of the theory is an essential
symmetry.  At lower energies non-relativistic quantum mechanical
models are powerful tools for understanding the dynamics of strongly
interacting nucleons.  At higher energies the physics is more
complicated, but one can expect that it is still dominated by a
manageable number of degrees of freedom.  Poincar\'e invariant quantum
mechanics is the only alternative to quantum field theory where it is
possible to realize the essential requirements of Poincar\'e
invariance,  spectral condition, and cluster properties
\cite{schroer:2007}.  It has the advantage that the Faddeev equation
provides a mathematically well-defined method for exactly solving the
strong interaction dynamics.  The Faddeev equation in this framework
is more complicated than the corresponding non-relativistic equation,
due to the non-linear relation between the mass and energy in
relativistic theories, but these difficulties can be overcome
\cite{Lin:2007ck,Keister:2005eq,kamada:2007}.  An
important advance that allows these calculations to be extended to
energies in the GeV range is the use of numerical methods based on
direct integrations, rather than partial wave expansions
\cite{Hang1,Hang2}.  These have been successfully applied to the
non-relativistic three-nucleon problem.  This paper demonstrates that they
can also be successfully applied to the relativistic problem, even
with its additional complications.
   
The model presented here involves three nucleons interacting with a
spin-independent Malfliet-Tjon \cite{malfliet} type of interaction.
It differs from more realistic interactions
\cite{AV18,CDBONN,NIJM} in that it is spin independent and
it does not give a high-precision fit to the two-body scattering
data.  In addition, the model is for fixed numbers of particles, not
allowing pion production, which is an open channel at these energies.
While these limitations must be addressed in realistic applications,
the three-body Faddeev calculations presented in this paper provide a
powerful framework for both testing approximations and for examining
the sensitivity of scattering observables to relativistic effects.
 
In order to investigate relativistic effects, we treat the interaction
as if it was determined by fitting the cross section obtained by
solving the non-relativistic Lippmann-Schwinger equation to scattering
data.  When this is done with a realistic interaction the experimental
differential cross section is properly transformed from the lab frame
to the center of momentum frame before the fit is done.  The result of
this process is that the computed differential cross section agrees
with the fully-relativistic experimental differential cross section in
the center of momentum frame as a function of the relative momentum.
Thus, even though the two-body scattering observables are computed
with a non-relativistic equation, there is nothing non-relativistic
about the result.  At the two-body level the corresponding
relativistic Lippmann-Schwinger equation must be designed to give the
same scattering observables.  This can be achieved by expressing the
relativistic mass operator as a simple function of the
non-relativistic center of momentum Hamiltonian
\cite{cps,bkrev}.  The important consequence of this is that it
does not make  sense to relate the relativistic and
non-relativistic two-body models using $p/m$ expansions; the
prediction of the relativistic and non-relativistic two-body models are
identical.  Real differences in the dynamics appear when the two-body
dynamical operators are used to formulate the three-body dynamics.
How this must be done in the two and three-body cases is dictated, up
to three-body interactions, by cluster properties.  The Faddeev
equation for the relativistic and non-relativistic system have
identical operators forms.  The permutation operators, two-body
transition operators and free resolvents that are input to the Faddeev
equation have different forms in the relativistic and non-relativistic
equations.  These differences are responsible for differences in the 
 relativistic and non-relativistic three-body calculations.
   
The calculations presented in this paper have a number of
consequences. The most important result is a demonstration that
direct integration methods can be successfully applied to extend the
energy range for converged solutions to Faddeev equations to
intermediate energies.  Our estimates of the number of partial waves
needed for calculations at different energies suggest that it is not
currently practical to extend existing partial wave calculations
beyond a few hundred MeV, while in this paper we have demonstrated
convergence of the direct integration methods for laboratory energies
up to 2 GeV.

While our model interaction is not realistic, when we compared the
results of our calculations to relativistic calculations at 200 MeV
that have been performed with realistic interactions
\cite{Witala1,Witala2,Witala3} in a partial wave basis, we found that
the qualitative features of the realistic model are reproduced in our
simple model, suggesting that some of the conclusions derived from our
model should be applicable to models with realistic
interactions.

Having a model where it is possible to perform numerically exact 
solutions of scattering observables in the intermediate energy 
range provides us with a tool to test approximations
that have been used in other calculations as well as to look for 
observables that are sensitive to the differences between the 
relativistic and non-relativistic models.

One common approximation that we tested is the replacement of
non-relativistic kinematic factors by the corresponding relativistic
kinematic factors in a non-relativistic model.  Our tests clearly
illustrated a big effect, but most of it is canceled by the associated
dynamical corrections.  This suggests that including only kinematic
corrections can actually provide large relativistic effects. Such  an approach 
should never be used in the absence of a complete theory where relativistic 
effects can be rigorously estimated.

A second important set of approximations are multiple scattering 
approximations.  These are expected to improve at higher energies,
but it is important to understand in the context of models based 
on realistic interactions how high these energies have to be
for convergence.

Our conclusion is that the convergence of the multiple scattering
series is non-uniform.  Even at 200~MeV our calculations show that the
first-order term reproduces the total elastic cross section; for the
total breakup cross section at least one more iteration is needed up
to about 600~MeV.  Both of these observations turn out to be
misleading when one investigates the differential cross sections.

While the total elastic cross section is reproduced at 200~MeV 
by the first order term,  the correct angular distribution requires
at least five orders in the multiple scattering series.  Even at 1~GeV   
the first order approximation is not accurate enough at forward angles.

For inclusive breakup reactions our computations show that the first
order calculation does not give the right size of the quasifree peak
even at 1~GeV, however for 800~MeV and above the second order term is
a good approximation.  For exclusive breakup the convergence of the
multiple scattering series even at 1~GeV energy depends on a specific
configuration.

Another type of approximation that is employed is to use on-shell
transition operators with a phenomenological representation of the
off-shell dependence.  In our formulation of the three-body problem that
off shell behavior needs to be computed by solving a singular integral
equation.  It was suggested in \cite{Keister:2005eq} that simply
replacing the off-shell two-body $T$ by its on-shell value might be a
good approximation.  This was based on the observation that the
difference between on and off shell Faddeev kernel is small.  Our
calculations show that while this does not lead to a large effect in
the elastic cross section, the off-shell effects lead to non-trivial
modifications when one considers the breakup cross sections.  This shows that
 such an approximation should not be used and also suggests that
phenomenological parameterizations of the off-shell behavior of the
two-body amplitudes need to be carefully tested, especially for
breakup reactions.

While a number of calculations have shown small relativistic effects
for the three-body binding energy, non-trivial effects have already
been observed in scattering observables at 200~MeV
\cite{Witala1,Witala2,Witala3}.  Our model confirms these previously
observed effects and indicates that they continue into the
intermediate energy region.  Our calculations exhibit a number of
sensitivities to relativistic effects in the breakup observables.
Both the shape and size of the quasielastic peak differ from the
non-relativistic quantities.
 
This paper demonstrates both the need for a relativistic description
of few-nucleon dynamics in the intermediate energy range and shows that the
problem is amenable to a numerically exact solution, using direct
integration, for laboratory energies up to 2~GeV.  In the future
relativistic few-body calculations will be important tools for testing
the validity of approximations, such as the Eikonal approximation.
Obviously extensions to include spin-dependent interactions, meson
channels, and interactions that are fit to higher energy data will be
needed for realistic applications.  The success of the calculations in
this paper provide a strong motivation for continuing this program.



\section*{Acknowledgments}
This work was performed in part under the
auspices of the U.~S.  Department of Energy, Office of Nuclear Physics,
under contract No. DE-FG02-93ER40756 with Ohio University,
contract No. DE-FG02-86ER40286 with the University of Iowa, and contract No.
DE-AC02-06CH11357 with Argonne National Laboratory. 
It was also partially supported by the Helmholtz Association through 
funds provided to the virtual institute ``Spin and strong QCD'' (VH-VI-231). 
We thank the Ohio  Supercomputer Center (OSC) for the use of
their facilities under grant PHS206. Part of the numerical calculations 
were performed on the IBM Regatta p690+ of the NIC in J\"ulich, Germany.


\clearpage

\begin{table}
\begin{tabular}{|c|c|c|} \hline
 $E_{lab}$ [GeV]
 &$\sigma_{op}$ [mb] $\quad \sigma_{tot}$ [mb] &$\sigma_{el}$ [mb] $\quad \sigma_{br}$
[mb]  \\
\hline
\hline
0.1  &349.4 \quad 350.6 &273.4 \quad 77.2 \\
\hline
0.2  &195.1 \quad 194.6 &158.6 \quad 36.0 \\
\hline
0.5  &106.2 \quad 106.8 &72.2 \quad 34.6 \\
\hline
0.8  &74.2  \quad 74.5  &46.6 \quad 27.9 \\
\hline
1.0  &62.3  \quad 61.8  &37.7 \quad 24.1 \\
\hline
1.2  &54.6  \quad 55.3  &33.0 \quad 22.3 \\
\hline
1.5  &43.7  \quad 44.9  &26.0 \quad 18.9 \\
\hline
2.0  &33.0  \quad 34.1  &18.9 \quad 15.2 \\
\hline
\end{tabular}
\caption{
The total elastic and break-up cross sections together with the total cross
section
extracted via the optical theorem calculated from a Malfliet-Tjon type potential 
as a function of the projectile laboratory kinetic energy. 
}
\label{table-1}
\end{table}

\vspace{5mm}

\begin{table}
\begin{tabular}{|c|c|c|c|c|} \hline
 $E_{lab}$ [GeV]   &q   $x_q$  $x^{q0}_{pq}$  $x_p$  p &q' x' $\phi$''
 &$\sigma_{op}$ [mb] \quad $\sigma_{tot}$ [mb] &$\sigma_{el}$ [mb] \quad
$\sigma_{br}$ [mb]  \\
\hline
1.0  &50  28  12  20  50   &50  20  20  &0.6154E+02 0.6069E+02
&0.3678E+02 0.2391E+02 \\
\hline
1.0  &50  32  12  20  50   &50  20  20  &0.6225E+02 0.6184E+02
&0.3774E+02 0.2411E+02 \\
\hline
1.0  &50  36  12  20  50   &50  20  20  &0.6257E+02 0.6233E+02
&0.3812E+02 0.2421E+02 \\
\hline
1.0  &50  40  12  20  50   &50  20  20  &0.6250E+02 0.6229E+02
&0.3809E+02 0.2420E+02 \\
\hline
\hline
1.0  &50  32  12  20  50   &50  20  20  &0.6225E+02 0.6184E+02
&0.3774E+02 0.2411E+02 \\
\hline
1.0  &50  32  12  20  50   &50  24  20  &0.6194E+02 0.6153E+02
&0.3753E+02 0.2400E+02 \\
\hline
1.0  &50  32  12  20  50   &50  28  20  &0.6199E+02 0.6133E+02
&0.3744E+02 0.2389E+02 \\
\hline
1.0  &50  32  12  20  50   &50  32  20  &0.6187E+02 0.6136E+02
&0.3746E+02 0.2390E+02 \\
\hline
\hline
1.0  &50  32  12  20  50   &50  20  20  &0.6225E+02 0.6184E+02
&0.3774E+02 0.2411E+02 \\
\hline
1.0  &50  32  16  20  50   &50  20  20  &0.6225E+02 0.6184E+02
&0.3773E+02 0.2411E+02 \\
\hline
1.0  &50  32  12  24  50   &50  20  20  &0.6225E+02 0.6177E+02
&0.3777E+02 0.2400E+02 \\
\hline
1.0  &50  32  12  20  50   &50  20  24  &0.6221E+02 0.6180E+02
&0.3773E+02 0.2408E+02 \\
\hline
\end{tabular}
\caption{\label{table-2}
The relativistic total elastic cross section, total breakup cross
section and total cross section extracted via the optical theorem calculated from
a Malfliet-Tjon type potential at 1~GeV as a function of the grid points. The
double prime quantities are the integration variables. 
}
\end{table}

\begin{table}
\begin{tabular}{|c|c|c|c|c|c|} \hline
 $E_{lab}$ [GeV]   &q   $x_q$  $x^{q0}_{pq}$  $x_p$  p &q' x' $\phi$'' &$\theta$
[deg] &$\frac{d\sigma}{d\Omega}$ [mb/sr]
 & $\Delta$ [\%] \\
\hline
1.0  &50  28  12  20  50   &50  20  20  &0.0  &0.6123E+03 &  \\
        &                     &            &21.8 &0.1142E+01 & \\
        &                     &            &62.1 &0.1159E-02 & \\
        &                     &            &102.3 &0.4193E-02 & \\
        &                     &            &151.5 &0.1233E-02 & \\
\hline
1.0  &50  32  12  20  50   &50  20  20  &0.0  &0.6266E+03 & 2.3 \\
        &                     &            &21.8 &0.1149E+01 & 0.6 \\
        &                     &            &62.1 &0.1268E-02 & 8.5 \\
        &                     &            &102.3 &0.4117E-02 & 1.9 \\
        &                     &            &151.5 &0.1233E-02 & 0.001 \\
\hline
1.0  &50  36  12  20  50   &50  20  20  &0.0  &0.6318E+03 & 0.8 \\
        &                     &            &21.8 &0.1170E+01 & 1.8 \\
        &                     &            &62.1 &0.1234E-02 & 2.8 \\
        &                     &            &102.3 &0.4127E-02 & 0.2 \\
        &                     &            &151.5 &0.1230E-02 & 0.2 \\
\hline
1.0  &50  40  12  20  50   &50  20  20  &0.0  &0.6319E+03 & 0.02  \\
        &                     &            &21.8 &0.1169E+01 & 0.1 \\
        &                     &            &62.1 &0.1234E-02 & 0.01 \\
        &                     &            &102.3 &0.4201E-02 & 1.8 \\
        &                     &            &151.5 &0.1234E-02 & 0.03 \\
\hline
\end{tabular}
\caption{\label{table-3}
The relativistic elastic differential cross sections for selected scattering angles
calculated at 1~GeV for a Malfliet-Tjon type potential as a function of 
the size of the $x_q$-grid.  
 The last column indicates
the percent difference with respect to the calculations for the corresponding
angle in the rows above.
}
\end{table}

\clearpage

\noindent

\vspace{5mm}

\end{document}